\documentclass[twocolumn,aps,amsmath,amssymb,prl]{revtex4-1}

\usepackage{amsmath}
\usepackage{amssymb}
\usepackage{bm}
\usepackage[usenames,dvipsnames]{color} 
\usepackage{cancel}
\usepackage{dcolumn}
\usepackage{epsf}
\usepackage{epsfig}
\usepackage{graphicx}
\usepackage{mathtools}
\usepackage{multirow}
\usepackage[final]{pdfpages}
\usepackage{setspace}
\usepackage{ulem} 
\usepackage{wrapfig}

\newcommand\Id{\leavevmode\hbox{\small1\normalsize\kern-.33em1}}
\newcommand{\half}{\frac{1}{2}}

\newcommand{\ket}[1]{\left\vert{#1}\right\rangle}

\newcommand{\ku}{\vert{0}\rangle}

\newcommand{\ave}[1]{\left\langle #1\right\rangle}

\newcommand{\ham}{{\mathcal{H}}}
\newcommand{\sx}{\sigma_x}
\newcommand{\sy}{\sigma_y}
\newcommand{\sz}{\sigma_z}

\definecolor{DarkBlue}{rgb}{0,0,.4}
 
\newcommand{\blist}[1]{
 \begin{list}{#1}
  { \setlength{\itemsep}{3pt}
     \setlength{\parsep}{2pt}
     \setlength{\topsep}{3pt}
     \setlength{\partopsep}{0pt}
     \setlength{\leftmargin}{1em}
     \setlength{\labelwidth}{1em}
     \setlength{\labelsep}{0.5em} } }
\newcommand{\elist}{
  \end{list}  }     

\DeclareMathSymbol{\vartheta}{\mathalpha}{letters}{"12}
\DeclareMathSymbol{\theta}{\mathalpha}{letters}{"23}
\DeclareMathSymbol{\phi}{\mathalpha}{letters}{"27}
\DeclareMathSymbol{\varphi}{\mathalpha}{letters}{"1E}

\definecolor{LinkColor}{rgb}{0,0,.5}
\usepackage[colorlinks=true,linkcolor=BrickRed,citecolor=LinkColor,urlcolor=LinkColor]{hyperref}
\renewcommand{\emph}{\textit}

\begin{document}

\title{Composite-pulse magnetometry with a solid-state quantum sensor}
\author{Clarice D.~Aiello}
\author{Masashi Hirose}
\author{Paola Cappellaro}
 \email{pcappell@mit.edu}
\affiliation{Department of Nuclear Science and Engineering, Massachusetts Institute of Technology, Cambridge, MA 02139, USA}

\date{\today}

\begin{abstract}
The sensitivity of quantum magnetometers~\cite{Taylor08} is challenged by control errors and, especially in the solid-state, by their short coherence times. 
Refocusing techniques can overcome these limitations and improve the sensitivity to periodic fields, but they come at the cost of reduced bandwidth and cannot be applied to sense static (DC) or aperiodic fields. 
Here we experimentally demonstrate that continuous driving of the sensor spin by a composite pulse~\cite{Levitt86} known as rotary-echo (RE)~\cite{Solomon57} yields a flexible magnetometry scheme, mitigating both driving power imperfections and decoherence. 
A suitable choice of RE parameters compensates for different scenarios of noise strength and origin. 
The method can be applied to nanoscale sensing in variable environments or to realize noise spectroscopy. 
In a room-temperature implementation based on a single electronic spin in diamond~\cite{Maze08,Balasubramanian08}, composite-pulse magnetometry provides a tunable trade-off between sensitivities in the $\mu$THz$^{-\frac{1}{2}}$ range, comparable to those obtained with Ramsey spectroscopy~\cite{Balasubramanian08}, and coherence times approaching $T_1$~\cite{Fedder11}. 

\end{abstract}
\maketitle

Solid-state quantum sensors attract much attention given their potential for high sensitivity and nano applications. 
In particular, the electronic spin of the nitrogen-vacancy (NV) color center in diamond is a robust quantum sensor due to a combination of highly desirable properties: optical initialization and read-out, long coherence times at room temperature ($T_1 >2$ ms~\cite{Jarmola12x,Waldherr12}, $T_2\gtrsim 0.5$ms \cite{Childress06}), the potential to harness the surrounding spin bath for memory and sensitivity enhancement~\cite{Cappellaro12b}, and bio-compatibility~\cite{McGuinness11}.

\begin{figure*}
\centering
\includegraphics[width=0.5\textwidth]{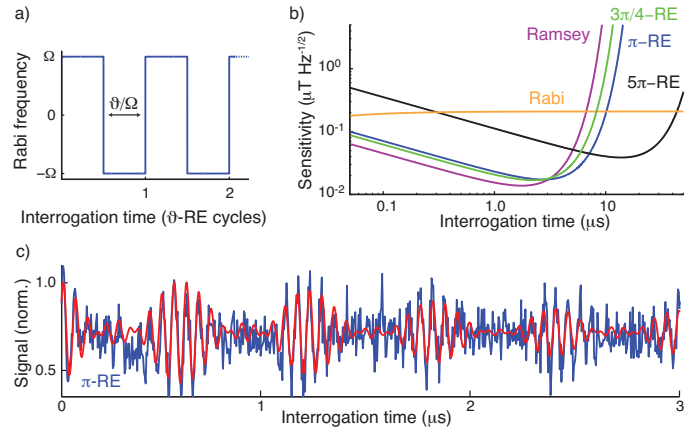}
\caption{\textbf{RE magnetometry scheme and sensitivity.} a) Experimental control sequence composed of $n$ RE, in which the phase of the microwave field is switched by $\pi$ at every pulse of duration $\frac{\theta}{\Omega}$. 
b) Magnetometry sensitivities $\eta_{RE}$ of $\theta = \{3\pi/4,\pi,5\pi\}$-RE sequences (green, blue, black), showing the tunability with the half-echo rotation angle. 
The sensitivity has its global minimum $\eta_{RE}\approx 1.38/\sqrt t$ (comparable to Ramsey magnetometry, purple) for $\theta \approx \frac{3\pi}{4}$ and consecutively increasing local minima for $\theta \approx (2k+1)\pi$. 
A decrease in sensitivity is followed by an increase in coherence time, which can approach $T_1$ as in Rabi-beat magnetometry (orange), whose sensitivity is limited by $\Omega$. 
Sensitivities are simulated in the presence of static bath noise using parameters from the fit depicted in c). 
c) A typical $n = 55$ cycles RE normalized fluorescence for $\theta = \pi$ and $\Omega\approx2\pi\!\times\!17$MHz (blue); the modulation in the signal is due to the hyperfine interaction with the $^{14}$N nucleus. 
The signal is filtered for even harmonics of $\pi\Omega/(\theta \mathrm{mod} 2\pi)$~\cite{SOM} and then fitted to Eq.~\ref{eq:RE} modified to include decoherence induced by static bath noise (red).}
\label{fig:RE}
\end{figure*}

Magnetometry schemes based on quantum spin probes (qubits) usually measure the detuning $\delta \omega$ from a known resonance. 
The most widely used  method is Ramsey spectroscopy~\cite{Ramsey90}, which measures the relative phase  $\delta \omega t$ the qubit acquires when evolving freely after preparation in a superposition state. 
 In the solid state, a severe drawback of this scheme is the short free-evolution dephasing time, $T_2^{\star}$, which limits the interrogation time. 
Dynamical decoupling (DD) techniques such as Hahn-echo~\cite{Hahn50} or CPMG~\cite{Meiboom58} sequences can  extend the coherence time. 
Unfortunately, such schemes also refocus the effects of static magnetic fields and are thus not applicable for DC-magnetometry. 
Even if $\delta \omega$ oscillates  with a known frequency (AC-magnetometry), DD schemes impose severe restrictions on the bandwidth, as the optimal sensitivity is reached only if the  field period matches the DD cycle time~\cite{Taylor08}.
Schemes based on continuous driving are thus of special interest for metrology in the solid-state because they can lead to extended coherence times~\cite{Kosugi05}. 
Recently, DC-magnetometry based on Rabi frequency beats was demonstrated~\cite{Fedder11}; in that method, a small detuning along the static magnetic field produces a shift $\approx \frac{\delta \omega^2}{2\Omega}$ of the bare Rabi frequency $\Omega$. 
Despite ideally allowing for  interrogation times approaching $T_1$, limiting factors such as noise in the driving field~\cite{Fedder11} and the bad scaling in $\delta \omega \ll \Omega$ make Rabi-beat magnetometry unattractive. 
More complex driving modulations~\cite{Levitt86} can provide not only a better refocusing of  driving field inhomogeneities, but also different scalings with $\delta\omega$, yielding improved magnetometry. 
In this work, we use a novel composite-pulse magnetometry method as a means of both extending coherence times as expected by continuous excitation, and keeping a good scaling with $\delta \omega$, which increases sensitivity. 
The $\theta$-RE is a simple composite pulse (Fig.~\ref{fig:RE}.a)  designed to correct for inhomogeneities in the excitation field; here, $\theta$ parametrizes the rotation angle of the half-echo pulse. 
For $\theta \neq 2\pi k$, $k \in \mathbb{Z}$, RE does not refocus magnetic fields along the qubit quantization axes and can therefore be used for DC-magnetometry. 
For $\theta=2\pi k$, RE provides superior decoupling from both dephasing~\cite{Laraoui11} and microwave noise and can be used to achieve AC-magnetometry.

In the rotating frame associated with the microwave field, and applying the rotating wave approximation, the Hamiltonian describing a continuous stream of $\theta$-REs is 
\[\mathcal{H}(t) = \frac{1}{2}\left[\Omega\, \mathbb{SW}(t)\sigma_x + \delta \omega (\Id - \sigma_z)\right],\] where $\mathbb{SW}(t)=\pm1$ is the square wave of period $T = \frac{2\theta}{\Omega}$. 
On resonance ($\delta \omega = 0$) the evolution is governed by the propagator $U_0 = e^{i \frac{\Omega}{2} \mathbb{TW}(t) \sigma_x}$, with $\mathbb{TW}(t)$ the triangular wave representing the integral of $\mathbb{SW}(t)$. 
We approximate the time evolution in the presence of a detuning $\delta\omega$ by a first order Average Hamiltonian expansion~\cite{Haeberlen76,SOM}, yielding an effective Hamiltonian over the cycle,  
\[\overline{\ham}^{(1)}=
  -\frac{\delta \omega}{\theta}\sin\left(\frac{\theta}{2}\right)\left[\cos\left(\frac{\theta}{2}\right)\sigma_z - \sin\left(\frac{\theta}{2}\right)\sigma_y\right].\]
Extending the above approximation to include the fast Rabi-like oscillations of frequency $\frac{\pi\Omega}{(\theta\mathrm{mod} 2\pi)}$,
we can thus calculate the population evolution for one of the qubit states, 
\begin{align}
\mathcal{S} & \approx  \frac{1}{2}+\frac{1}{2}\cos^2\left(\frac{\theta}{2}\right) + \frac{1}{2}\sin^2\left(\frac{\theta}{2}\right) \times \nonumber \\
&  \times \cos\left(\frac{2\delta\omega t}{\theta}\sin\left(\frac{\theta}{2}\right)\right)\cos\left(\frac{\pi\Omega t}{(\theta    \mathrm{mod} 2\pi)}\right).
\label{eq:RE}
\end{align} 
The signal $\mathcal{S}$ indicates the presence of two spectral lines at $\frac{\pi\Omega}{(\theta    \mathrm{mod} 2\pi)} \pm \frac{2\delta \omega}{\theta}\sin\left(\frac{\theta}{2}\right)$ for a detuning $\delta \omega$. 
 
Thanks to this linear dependence on $\delta \omega$, we expect a favorable scaling of the sensitivity $\eta$,  given by the  shot-noise-limited magnetic field resolution per unit measurement time~\cite{Wineland92, Taylor08}. 
For $N$ measurements and a signal standard deviation $\Delta{\mathcal{S}}$, the sensitivity is
\begin{equation}
\eta =\Delta B \sqrt{\mathcal{T}} =  \frac{1}{\gamma_e}\lim_{\delta \omega \rightarrow 0}\frac{\Delta\mathcal{S}}{|\frac{\partial \mathcal{S}}{\partial \delta\omega}|} \sqrt{{N} (t + t_d)}\ ,
\end{equation}
where $\gamma_e$ ($\approx 2.8$MHzG$^{-1}$ for NV) is the sensor gyromagnetic ratio and {$\Delta B$ is the minimum detectable field.} 
We broke down the total measurement time $\mathcal{T}$  into interrogation time $t$ and  dead-time $t_d$ required for initialization and readout. 
In the absence of relaxation, and neglecting $t_d$, a RE magnetometer interrogated at complete echo cycles $t = n\frac{2\theta}{\Omega}$ yields $\eta_{RE} = \frac{1}{\gamma_e\sqrt{t}}\frac{\theta}{2\sin^2(\theta/2)}$. 
As shown in Fig.~\ref{fig:RE}.b, RE magnetometry has thus sensitivities comparable to Ramsey spectroscopy, $\eta_{Ram} \approx \frac{1}{\gamma_e\sqrt{t}}$. 
Conversely, Rabi-beat magnetometry has $\eta_{Rabi} \approx \frac{\sqrt{2\Omega}}{\gamma_e}$ at large times~\cite{SOM}, which makes it unsuitable for magnetometry despite long coherence times.

\begin{figure*}
\centering
\includegraphics[width=0.4\textwidth]{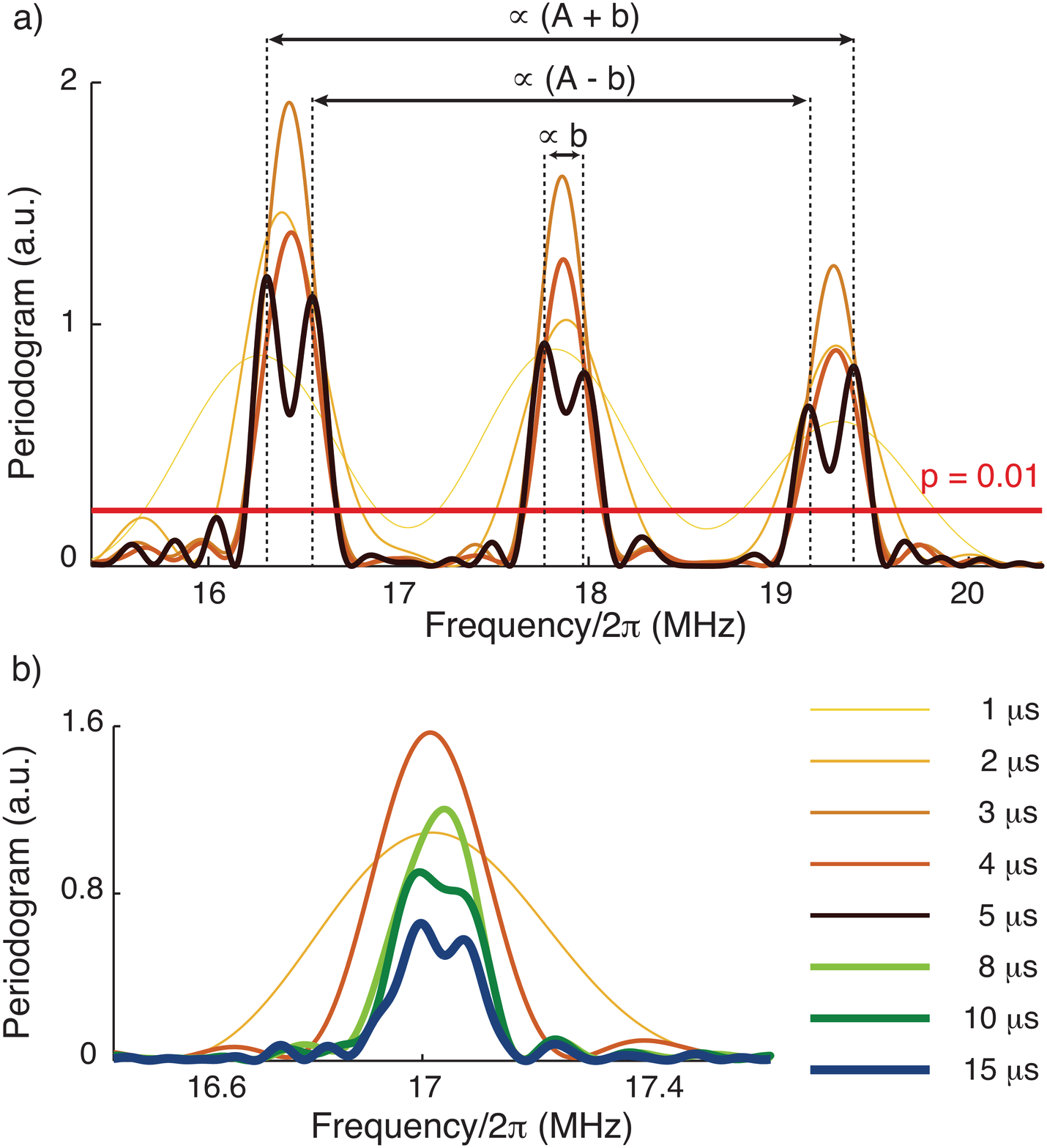}
\caption{\textbf{The periodogram identifies the frequency content of the signal.} a) Periodogram for $\pi$-RE sequence for increasing interrogation times (thicker lines). 
The periodogram is defined as the squared magnitude of the Fourier transform of the time signal. 
A pair of symmetric peaks about $\Omega$ signals the existence of one detuning $\delta \omega$. 
The number of resolved frequencies increases with time, at the expense of signal-to-noise ratio. 
After $5\mu$s of interrogation, we can estimate both the hyperfine interaction $A\approx 2\pi\!\times\!(2.14\pm0.03)$MHz and a small residual detuning from the presumed resonance, $b\approx 2\pi\!\times\!(0.17\pm 0.02)$MHz. 
In this estimate, we correct for the real rotation angle $\theta\approx 0.96\pi$ using the difference between the nominal and experimentally realized Rabi frequency (symmetry point in the spectrum). 
The uncertainty in the measurement is estimated taking into account the total interrogation time, the number of points in the time-domain signal, and the $S/N$, as detailed in~\cite{SOM}. 
Periodogram peaks can be tested for their statistical significance~\cite{Wei06,SOM}; we confirm that all 6 frequency peaks are considerably more significant than a $p = 0.01$ significance level (red). 
b) Innermost pair of frequency peaks arising from a $b\approx 2\pi\!\times\!(64\pm12)$kHz residual detuning in another experimental realization, for an interrogation time of $15\mu$s.}
\label{fig:Periodogram}
\end{figure*}

To establish the sensitivity limits of RE magnetometry and compare them to other DC-magnetometry strategies, we carried out proof-of-principle experiments in single  NV centers in a bulk electronic-grade diamond sample. 
A  static magnetic field $B_{\parallel}\approx100$ G effectively singles out a qubit $\{\ket{0},\ket{1}\}$ from the NV ground-state spin triplet, as the Zeeman shift lifts the degeneracy between the $\ket{\pm1}$ levels. 
The qubit is coupled to the spin-1 $^{14}$N nucleus that composes the defect by an isotropic hyperfine interaction of strength  $A\approx2\pi\!\times\!2.17$MHz. 
After optical polarization into state $|0\rangle$, we apply a stream of $n$  RE cycles using microwaves with frequency $\omega$  close  to the qubit resonance  $\omega_0 = \Delta+\gamma_eB_\parallel$, where $\Delta=2.87$GHz is the NV zero-field splitting. 
Because of the hyperfine coupling, $\omega_0$ is the resonant frequency only when the nuclear state is $m_I = 0$. 
At room temperature, the nitrogen nucleus is unpolarized and, while its state does not change over one experimental run, in the course of the ${N}\sim10^{6}$ experimental realizations, $\approx\!2/3$  of the times the qubit is off-resonantly driven by $|\delta\omega|  =  A$. 
A typical $n$-cycle RE fluorescence signal is plotted in Fig.~\ref{fig:RE}.c for $\theta = \pi$ and $\Omega\approx2\pi\!\times\!17$MHz, while  to determine the frequency content of the signal we plot the periodogram in Fig.~\ref{fig:Periodogram}.

The number of distinguishable frequencies increases with interrogation time at the expense of signal-to-noise ratio. 
RE-magnetometry not only discriminates the frequency shifts due to \mbox{$A\approx2\pi\!\times\!(2.14\pm0.03)$MHz} but, for interrogation times as short as $5\mu$s, also those due to a small residual detuning \mbox{$b\approx2\pi\!\times\!(0.17\pm0.02)$MHz} from the presumed resonance. 
In contrast, under the same experimental conditions Rabi magnetometry does not discern such a detuning before an interrogation time $\approx 188\mu$s~\cite{SOM}. 
With  longer interrogation times $\sim\!15\mu$s as in Fig.~\ref{fig:Periodogram}.b, we can detect a frequency as small as $b \approx 2 \pi\!\times\!(64 \pm 12)$~kHz.

\begin{figure*}
\includegraphics[width=\textwidth]{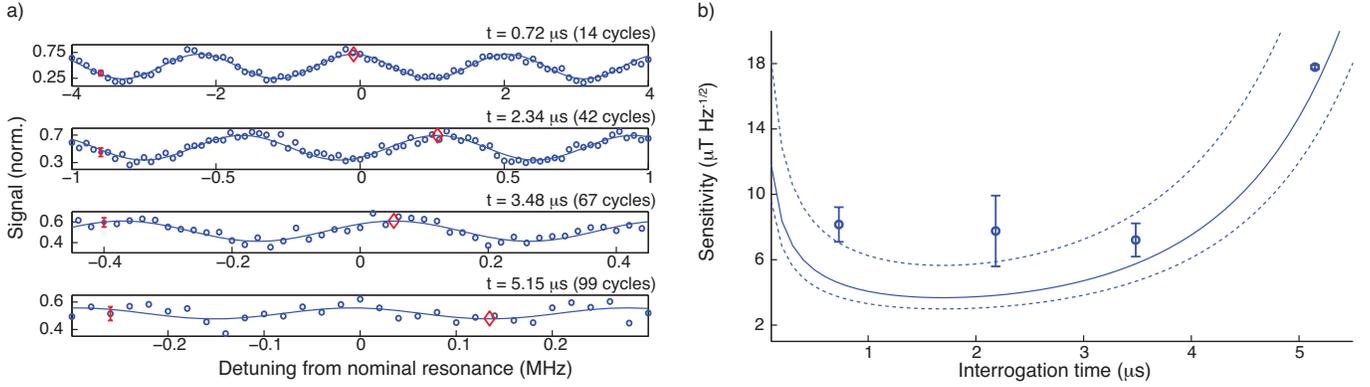}
\caption{\textbf{Experimental sensitivity of RE magnetometry.} a) RE signals at fixed interrogation times as a function of the detuning $\delta\omega$ from  resonance, from which $|\frac{\partial \mathcal{S}}{\partial \delta\omega}|$ is numerically calculated to obtain the sensitivity $\eta$. 
With increasing interrogation times, the slopes initially increase, indicating an improvement in $\eta$; the effect of decoherence for the longer interrogation times degrades the sensitivity, and the slopes smoothen accordingly. 
The different amplitude modulations are due to the three-frequencies in the signal, $\{b, A \pm b\}$; polarizing the nuclear spin~\cite{Jacques09} would eliminate this modulation. 
From the fitted resonances for each curve (red diamonds), we estimate the true resonance to be at $0.09 \pm 0.15$MHz from the presumed resonance. 
Typical errors in the measurement are indicated (red errorbars). 
Interrogation times are chosen to coincide with minima of the sensitivity in the presence of the hyperfine interaction; in other words, $C_A$ is at a local maximum at those times~\cite{SOM}. 
b) For each fixed interrogation time, the minimum sensitivity $\eta$ within one oscillation period of the fitted oscillation frequency obtained in a), $\tau = \frac{2t\sin(\theta/2)}{\theta}$, is plotted. 
The experimental points agree in trend with the theoretically expected sensitivities $\eta/(C\!\times\!C_A)$ (solid curves), here corrected for the presence of static bath noise, $\eta\to\eta e^{(t/T'_{RE})^2}$. 
$T'_{RE}$ was computed using a $T_2^{\star} \approx2.19\pm0.15\mu$s fitting from a Ramsey decay experiment~\cite{SOM}.
}
\label{fig:Sens}
\end{figure*}

To determine the experimental sensitivities, we estimate $|\frac{\partial \mathcal{S}}{\partial \delta\omega}|$ by driving the qubit with varying $\omega$, at fixed interrogation times $t$ (Fig.~\ref{fig:Sens}.a).
For each $t$, in Fig.~\ref{fig:Sens}.b we plot the minimum $\frac{1}{\gamma_e}\frac{\Delta\mathcal{S}}{|\frac{\partial \mathcal{S}}{\partial \delta\omega}|}\sqrt{N t}$ and compare it to the adjusted theoretical sensitivity $\eta/(C\!\times\!C_A)$. 
Here $(C\!\times\!C_A)$, $\approx (5.9\pm1.4) \times 10^{-3}$ in our setup, is a factor taking into account readout inefficiencies and a correction for the presence of the hyperfine interaction~\cite{Taylor08,SOM}. 
The sensitivities agree with the theoretical model, with optimal  $\sim 10\mu$THz$^{-\frac{1}{2}}$. 
Improved sensitivities are expected from isotopically pure diamond~\cite{Balasubramanian09}; adequate choice of interrogation times or polarization of the nuclear spin can easily set $C_A=1$, while $C$ can be improved by efficient photon collection~\cite{Babinec10} or  using repeated readout methods~\cite{Neumann10b,SOM}.

The sensitivity of a NV magnetometer is ultimately limited by the interaction of the quantum probe with the nuclear spin bath.
We  model the effect of the  spin bath by a classical noise source along $\sigma_z$~\cite{Klauder62}, described by an Ornstein-Ulhenbeck (OU) process of strength $\sigma$ and correlation time $\tau_c$. 
In the limit of long $\tau_c$ (static bath), the dephasing time associated with RE (Ramsey) magnetometry is $T'_{RE} = \frac{\theta}{\sigma\sqrt{2}|\sin(\theta/2)|}$ ($T_2^{\star} = T'_{Ram} = \frac{\sqrt{2}}{\sigma}$) respectively (see~\cite{SOM} for the general case). 
While at the optimum interrogation time $\frac{T'}{2}$ one has $\eta_{RE}/\eta_{Ram} =\sqrt{\frac{\theta}{2\sin(\theta/2)^3}} > 1$, RE magnetometry allows a greater flexibility in choosing the effective coherence time, as larger $\theta$ increase the resilience  to  bath noise. 
Thus one can match the RE interrogation time to the duration of the field one wants to measure. 
In addition, RE can yield an overall advantage when taking into consideration the dead-time $t_d$. 
If $t_d\gg T'_{Ram}$, as in repeated readout methods~\cite{Neumann10b}, a gain in sensitivity can be reached by exploiting the longer interrogation times enabled by RE magnetometry~\cite{SOM}. 
An even larger advantage is given by AC-magnetometry with $2\pi k$-RE~\cite{Hirose12}, since RE provides better protection than pulsed DD schemes~\cite{Laraoui11,Hirose12}.

\begin{figure*}[ht]
\centering
\includegraphics[width=0.47\textwidth]{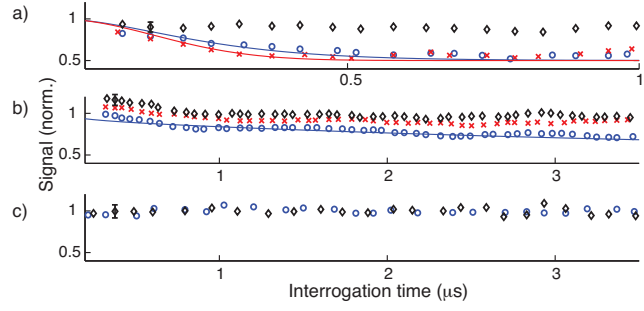}
\caption{\textbf{Experimental implementation of microwave frequency noise, against which RE is robust.} 40 realizations of both static (red) and OU ($\tau_c = 200$ns, blue) microwave noise of strength 0.05$\Omega$, $\Omega \approx 2\pi\!\times\!19$MHz, for a) Rabi, b) $5\pi$-RE and c) $\pi$-RE sequences. We note that the correlation time of the driving field noise was purposely set much shorter than what usually observed in experiments in order to enhance the different behaviour of $\pi$-RE and $5\pi$-RE.
Typical errors in the measurement are indicated (black errorbars). 
a) We plot the peak of the Rabi fringes in the presence of static (red crosses) and OU (blue circles) microwave noise, which are in agreement with the expected theoretical decay detailed in~\cite{SOM} (solid blue line and solid red line, respectively). 
The oscillations at the tail of the signals are due to the finite number of experimental realizations. 
The peaks of a no-noise Rabi experiment are plotted for comparison (black diamonds). 
b) The peak of the $5\pi$-RE revivals are plotted in the presence of static (red crosses) and OU (blue circles) microwave noise; the peaks of the $5\pi$-RE in the absence of microwave noise are also plotted for comparison (black diamonds). 
Whereas the echo virtually does not decay in the presence of static noise, under the effect of stochastic noise, the echo decays only weakly as stipulated by theory (solid blue line) described in~\cite{SOM}. 
c) The peak of the $\pi$-RE revivals are plotted in the presence of OU microwave noise (blue circles); virtually no decay is found if compared to the peaks of a no-noise experiment (black diamonds).}
\label{fig:Noise}
\end{figure*}

Excitation field instabilities along $\sigma_x$ also accelerate the decay of RE and Rabi signals. 
However, provided the echo period is shorter than $\tau_c$, RE magnetometry corrects for stochastic noise in Rabi frequency~\cite{SOM}. 
This protection was demonstrated experimentally by applying static and OU noise ($\tau_c\approx 200$ns)  in the excitation microwave, both with  strength $0.05\Omega$. 
The results for Rabi and $\{\pi, 5\pi\}$-RE sequences in Fig.~\ref{fig:Noise} clearly show that whereas the Rabi signal decays within $\approx 0.5\mu$s, $5\pi$-RE refocuses static excitation noise and presents only a very weak decay under finite-correlation noise after much longer interrogation times $\approx 3\mu$, in agreement with theory detailed in~\cite{SOM}; $\pi$-RE is robust against the same noise profiles.

The unique ability of the RE magnetometer to adjust its response to distinct noise sources is relevant when the sample producing the magnetic field of interest is immersed in a realistic environment; moreover, the field source might itself have a finite duration or duty cycle.
The advantage is two-fold: the protection from noise can be tuned by changing the echo angle, thus allowing  the interrogation times to be varied. 
Techniques for repeated readout in the presence of a strong  magnetic field $\gtrsim 1000$G~\cite{Neumann10b,SOM} can at once improve sensitivities and enable the use of much lower qubit resonance frequencies $\sim$MHz, preferable in biological settings. 
Additionally, a RE-magnetometer can discriminate magnetic noise sources given the sensor's well-understood decoherence behavior under different noise profiles, effectively enabling noise spectroscopy for both $\sigma_z$ and $\sigma_x$-type noises.

NV center-based RE magnetometry could find useful application, for example, to sense the  activity of differently-sized calcium signaling domains in living cells, more specifically in neurons. 
Transient calcium fluxes regulate a myriad of cell reactions~\cite{Augustine03}. 
The signaling specificity of such fluxes is determined by their duration and mean travelled distance between membrane channel and cytoplasm receptor. 
The smaller, faster-signaling domains have resisted thorough investigation via both diffraction-limited optical microscopy~\cite{Augustine03}, and the use of fluorescing dyes, which do not respond fast or accurately enough to Ca$^{2+}$ transients~\cite{Keller08}. 
The magnetic field produced by as few as $10^5$ Ca$^{2+}$, being diffused within $\sim 10\mu$s through a $\sim 200$nm domain, can be picked up by a nanodiamond scanning sensor~\cite{Degen08,Grinolds11} with sensitivity $\sim$ $10\mu$THz$^{-\frac{1}{2}}$ placed at close proximity ($\sim 10$nm)~\cite{SOM}. 
The trade-off between sensitivity and optimal interrogation time under RE magnetometry can be optimized to the characteristics of the signaling domain under study by a suitable choice of $\theta$.

In conclusion, we have demonstrated a quantum magnetometry scheme based on composite-pulses. 
Its key interest stems both from the continuous-excitation character, offering superior performance for solid state sensors such as the NV center, and from the possibility of tuning the sensor's coherence time and sensitivity in the presence of variable or unknown sensing environments, to protect from or map noise sources. 
Current technology enables immediate implementation of such scheme at the nanoscale.

\section*{Methods}
The NV center is a naturally occurring point defect in diamond, composed of a vacancy adjacent to a substitutional nitrogen in the carbon lattice. 
The ground state of the negatively charged NV center is a spin-triplet with zero-field splitting $\Delta=2.87$GHz between the $m_{s}=0$ and $m_{s}=\pm1$ sub-levels. 
Coherent optical excitation at $532$nm promotes the quantum state of the defect non-resonantly to the first orbital excited state. 
While the $m_{s}=0$ state mostly relaxes with phonon-mediated fluorescent emission ($\approx650\!-\!800$nm), the  $m_{s}=\pm1$ states have in addition an alternative, non-radiative decay mode to the $m_s=0$ state via  metastable singlet states. 
Due to this property, each ground state is distinguishable by monitoring the intensity of emitted photons during a short pulse of optical excitation. 
Additionally, continuous optical excitation polarizes the NV into the $m_s=0$ state. 
We apply a magnetic field ($\approx 100$G) along a crystal axis $\ave{111}$ to lift the degeneracy between the $m_{s}=\pm1$ states and drive an effective two-level system $m_s=\{0,1\}$ at the resonant frequency ($\omega_0\approx 3.15$GHz) obtained by continuous wave electron spin resonance and Ramsey fringe experiments. 
In each  experimental run, we normalize the signal with respect to the reference counts from the $m_{s}=\{0,1\}$ states, where the transfer to state $m_s= 1$ is done by adiabatic passage. 
An arbitrary waveform generator is used to shape the control microwave field, which is delivered by a stripline in contact with the sample. 
A detailed setup description is found in~\cite{SOM}.

\section*{Acknowledgements}
C.~D.~A. 
thanks Boerge Hemmerling, Michael G.~Schmidt and Marcos Coque Jr.~for stimulating discussions. 
This work was supported in part by  the
U.S. 
Army Research Office through a MURI grant No. 
W911NF-11-1-0400 and by DARPA (QuASAR program). 
C.~D.~A. 
acknowledges support from the Schlumberger Foundation.
%
%


\newpage ~ \newpage ~ 

\newpage ~ \newpage ~ 

\newpage ~ \newpage ~ 

\newpage ~ \newpage ~ 

\pagebreak

\onecolumngrid

\begin{center}
\large{\textbf{Supplement to ``Composite-pulse magnetometry with a solid-state quantum sensor''}}
\end{center}

\tableofcontents

\section{\label{Setup}Setup}

Experiments were run at room-temperature with  single NV centers from an electronic grade single crystal plate ({[}100{]} orientation, Element 6) with a substitutional nitrogen concentration $<5$ ppb.
The fluorescence of single NV centers is identified by a home-built confocal scanning microscope. The sample is mounted on a piezo stage (\mbox{Nano-3D200}, Mad City Labs).  The excitation at $532$nm is provided by a  diode-pumped laser (Coherent Compass 315M), and fluorescence in the phonon sideband ($\sim 650-800$nm) is collected by a $100$x, \mbox{NA $=1.3$} oil immersion objective (Nikon Plan Fluor). The fluorescence photons are collected into a single-mode broadband fiber of \mbox{NA $=0.12$} (Font Canada) and sent to a single-photon counting module (\mbox{SPCM-AQRH-13-FC}, Perkin Elmer) {with  acquisition time 100 or 200 ns.}

Laser pulses for polarization and detection are generated by an acousto-optic modulator with rise time $\lesssim 7$ns (\mbox{1250C-848}, Isomet). A signal generator (\mbox{N5183A-520}, Agilent) provides microwave fields to coherently manipulate the qubit. An arbitrary waveform generator at $1.2$GS/s (AWG5014B, Tektronix) is employed to shape microwave pulses with the help of an I/Q mixer (\mbox{IQ-0318L}, Marki Microwave), and to time the whole experimental sequence. Microwaves are amplified (\mbox{GT-1000A}, Gigatronics) and subsequently delivered to the sample by a copper microstrip mounted on a printed circuit board, fabricated in MACOR to reduce losses. 

A static magnetic field is applied by a permanent magnet (\mbox{BX0X0X0-N52}, K$\&$J Magnetics) mounted on a rotation stage, which at its turn is attached to a three-axis translation stage; this arrangement enables the adjustment of the magnetic field angle with respect to the sample. The magnetic field is  aligned along a {[}111{]} axis by maximizing the Zeeman splitting in a CW ESR spectrum. 

\section{\label{Dyn}Dynamics under rotary echo sequence}

Consider a two-level system, $|0\rangle$ and $|1\rangle$, with resonance frequency $\omega_0$. The qubit is excited by radiation of frequency $\omega$ with associated Rabi frequency $\Omega$ and phase modulation $\phi(t)$, such that the magnetic field amplitude is $\Omega\cos(\omega t + \phi(t))$.
The Hamiltonian is then
\begin{equation}
\mathcal{H}_{lab} =  \left( \begin{array}{cc}
0 & \Omega\cos(\omega t + \phi(t)) \\
\Omega\cos(\omega t + \phi(t)) & \omega_{0} \end{array} \right) \,.
\end{equation}
In a frame  rotating with the excitation field, the operator 
\begin{equation}
U_{rot} = \left( \begin{array}{cc}
1 & 0 \\
0 & e^{i\omega t} \end{array} \right)
\end{equation}
transforms the Hamiltonian to
\begin{equation}
\mathcal{H} 
 =  \left( \begin{array}{cc}
0 & \Omega\cos(\omega t + \phi(t))e^{-i\omega t} \\
\Omega\cos(\omega t + \phi(t))e^{i\omega t} & \omega_{0} - \omega \end{array} \right) .
\end{equation}
Applying the rotating wave approximation and setting set $\delta\omega \equiv \omega_{0} - \omega$, the Hamiltonian reads
\begin{equation}
\mathcal{H} \approx \left( \begin{array}{cc}
0 & \frac{\Omega}{2} e^{i\phi(t)} \\
\frac{\Omega}{2} e^{-i\phi(t)} & \delta\omega \end{array} \right).
\end{equation}

One rotary echo (RE) is composed of two identical pulses of nominal rotation angle $\theta$ applied with excitation phases shifted by $\pi$. Under a sequence of RE, $e^{i\phi(t)} = \mathbb{SW}(t)$, with $\mathbb{SW}(t)$ the square wave of period $T = \frac{2\theta}{\Omega}$ equal to the RE cycle time:
\begin{equation}
\mathbb{SW}(t) = \frac{4}{\pi}\sum_{k=1,\text{odd}}^{\infty}\frac{1}{k}\sin\left(\frac{k\pi\Omega t}{\theta}\right).
\end{equation}
On resonance ($\delta\omega = 0$) the evolution operator is trivially obtained:
\begin{equation}
U_0 =  \left( \begin{array}{cc}
\cos(\frac{\Omega}{2}\mathbb{TW}(t)) & -i\sin(\frac{\Omega}{2}\mathbb{TW}(t)) \\
 -i\sin(\frac{\Omega}{2}\mathbb{TW}(t))& \cos(\frac{\Omega}{2}\mathbb{TW}(t)) \end{array} \right)=\cos\left(\frac{\Omega}{2}\mathbb{TW}(t)\right)\Id-i\sin\left(\frac{\Omega}{2}\mathbb{TW}(t)\right)\sx,
\end{equation}
where $\mathbb{TW}(t)$ is the triangular wave representing the integral of $\mathbb{SW}(t)$,
\begin{equation}
\mathbb{TW}(t) = \frac{\theta}{2\Omega} - \frac{4\theta}{\pi^2\Omega}\sum_{k=1,\text{odd}}^{\infty}\frac{1}{k^2}\cos\left(\frac{k\pi\Omega t}{\theta}\right) \ .
\end{equation}

Using $U_0$ we make a transformation to the toggling frame of the microwave~\cite{Duer04}  to obtain the Hamiltonian $\tilde{\ham}$:
\begin{equation}
\tilde{\mathcal{H}}
=  \frac{\delta\omega}{2}\left( \begin{array}{cc}
1-\cos(\Omega\mathbb{TW}(t)) & i\sin(\Omega\mathbb{TW}(t)) \\
-i\sin(\Omega\mathbb{TW}(t)) & 1+\cos(\Omega\mathbb{TW}(t)) \end{array} \right)=\frac{\delta\omega}2\left[\Id-\cos(\Omega\mathbb{TW}(t))\sz-\sin(\Omega\mathbb{TW}(t))\sy\right].
\end{equation}
$\tilde{\mathcal{H}}$ is periodic with $T$ and has a strength $T\delta\omega \ll 1$, and can thus be analyzed with an average Hamiltonian expansion. In order to do so, we first express the elements of $\tilde{\mathcal{H}}$ in their Fourier series:
\begin{equation}
\cos(\Omega\mathbb{TW}(t)) = \frac{\sin\theta}{\theta} + 2\theta\sin\theta\sum_{k=1,\text{odd}}^{\infty}\frac{(-1)^k}{\theta^2 - k^2\pi^2}\cos\left(\frac{k\pi\Omega t}{\theta}\right);
\end{equation}
\begin{equation}
\sin(\Omega\mathbb{TW}(t)) = \frac{1-\cos\theta}{\theta} + 2\theta\sum_{k=1,\text{odd}}^{\infty}\frac{(-1)^k((-1)^k -\cos\theta)}{\theta^2 - k^2\pi^2}\cos\left(\frac{k\pi\Omega t}{\theta}\right) \ .
\end{equation}
To first order then,
\begin{equation}
\mathcal{\overline{H}}^{(1)} = \frac{1}{T}\int_{0}^{T}\mathcal{\tilde{H}}(t')\mathrm{d}t'=\frac{\delta\omega}{\theta}\sin\!\left(\frac{\theta}{2}\right)\!\left( \begin{array}{cc}
-\cos\left(\theta/2\right) & i\sin\left(\theta/2\right) \\
-i\sin\left(\theta/2\right)& \cos\left(\theta/2\right)\end{array} \right)=-\frac{\delta\omega}{\theta}\sin\!\left(\frac{\theta}{2}\right)\!\left[\cos\!\left(\frac\theta2\right) \sz-\sin\!\left(\frac\theta2\right)\sy\right].
\end{equation}
For $n$ rotary cycles, the propagator is approximated by $U_{RE} = e^{i\mathcal{\tilde{H}}(t)t} \approx e^{inT\mathcal{\overline{H}}^{(1)}}$. The population of a system initially prepared in $|0\rangle$, under the action of $U_{RE}$, is described at full echo times by the signal
\begin{equation}
\mathcal{S}(n) \approx \frac{1}{2}\left[1 + \cos^2\left(\frac{\theta}{2}\right) + \sin^2\left(\frac{\theta}{2}\right)\cos\left(\frac{4\delta\omega n}{\Omega}\sin\left(\frac{\theta}{2}\right)\right)\right] \ .
\end{equation}
Extending the above approximation to include the fast Rabi-like oscillations of frequency $\frac{\pi\Omega}{(\theta \mathrm{mod} 2\pi)}$, we obtain 
\begin{equation}
\mathcal{S}(t)  \approx  \frac{1}{2}\left[1 +\cos^2\left(\frac{\theta}{2}\right) + \sin^2\left(\frac{\theta}{2}\right)  
\cos\left(\frac{2\delta\omega t}{\theta}\sin\left(\frac{\theta}{2}\right)\right)\cos\left(\frac{\pi\Omega t}{(\theta  \mathrm{mod} 2\pi)}\right)\right] \ ,
\end{equation}
indicating the presence of two spectral lines at $\frac{\pi\Omega}{(\theta    \mathrm{mod} 2\pi)} \pm \frac{2\delta \omega}{\theta}\sin\left(\frac{\theta}{2}\right)$ for each existing detuning $\delta \omega$. 

Our numerical studies suggest the existence of further signal components arising from higher frequency components in the Fourier expansion, which are not contemplated by the first-order approximation outlined above. Such components are $\propto \cos\left(\frac{2\delta\omega t}{\theta}\sin\left(\frac{\theta}{2}\right)\right)\cos\left(\frac{(2k+1)\pi\Omega t}{(\theta \mathrm{mod} 2\pi)}\right)$ and $\propto \cos\left( \frac{2k\pi\Omega t}{(\theta  \mathrm{mod} 2\pi)}\right)$, $k \in \mathbb{Z}$, thus being linked to split pairs of spectral lines around $\frac{(2k+1)\pi\Omega t}{(\theta \mathrm{mod} 2\pi)}$, and single lines at $\frac{2k\pi\Omega t}{(\theta  \mathrm{mod} 2\pi)}$. 

\section{\label{Periodo}Periodogram}

The periodogram is defined as the squared magnitude of the Fourier transform (FT) of the signal $\mathcal{S}(t)$ at times $t_j$ ($j=1,\dots,M$), $\mathcal P\equiv\frac{1}{M}|\sum_{j=1}^{M}d_j e^{i\omega t_j}|^2$, where $d_j$ are the $M$ data points~\cite{Bretthorst88}. 

Unlike the FT, the periodogram does provide bounds for frequency estimation from spectral analysis, besides being able to accommodate for  noise profiles beyond static and white noise~\cite{Bretthorst88}. Take a simple sinusoidal signal $d_t = K\cos(2\pi f t) + e_t$, where $e_t$ is the added noise characterized by a (least informative) Gaussian probability distribution \textsf{Normal}(0,$\sigma^2$), with $\sigma$ in circular frequency units. To $\sigma$ is assigned Jeffrey's prior $\frac{1}{\sigma}$, which indicates complete ignorance of this scale parameter. Under these conditions, the estimate frequency content of the signal is given by $f_{est} = f_{peak} \pm \delta f$, where $f_{peak}$ is the frequency of the periodogram peak, and 
\begin{equation}
\delta f = \frac{2\sqrt{3}}{\pi}\frac{\sigma}{K t \sqrt{M}} \ . 
\end{equation}
Here $t$ is the total interrogation time for the $M$ data points. $\delta f$ correctly takes into account the effect of both the interrogation duration $t$ and the $S/N \equiv \frac{K_{\text{RMS}}}{\sigma} = \frac{K}{\sqrt{2}\sigma}$, and is shown to correspond to the classical Cramer-Rao bound~\cite{Rife74}. Note that $\delta f$ is in general smaller than the so-called Fourier limit, $\delta f_{Fl} = \frac{1}{2t}$. The method is readily applicable to signals with multiple frequency content $\{f_i\}$. 

In Supplementary Fig.~\ref{fig:SOMcompPer}, we compare typical experimental periodograms for $\pi$-RE, Ramsey and Rabi signals taken under the same conditions, for increasing interrogation times. The Ramsey periodogram, despite its lower signal intensity, clearly shows the 3 detunings $\{b+2\pi\!\times\!5\text{MHz}, A \pm (b+2\pi\!\times\!5\text{MHz})\}$ present in the signal after  $1\mu$s; $\pi$-RE is sensitive to the residual detuning $b \sim 2\pi\!\times\!0.17$MHz as explained in the main text after $5\mu$s; finally, the Rabi sequence would only become sensitive to $b$ after an interrogation time $\sim 188\mu$s, which is reflected in the broad single peak of the periodogram.

\begin{figure}
\centering
\includegraphics[width=\textwidth]{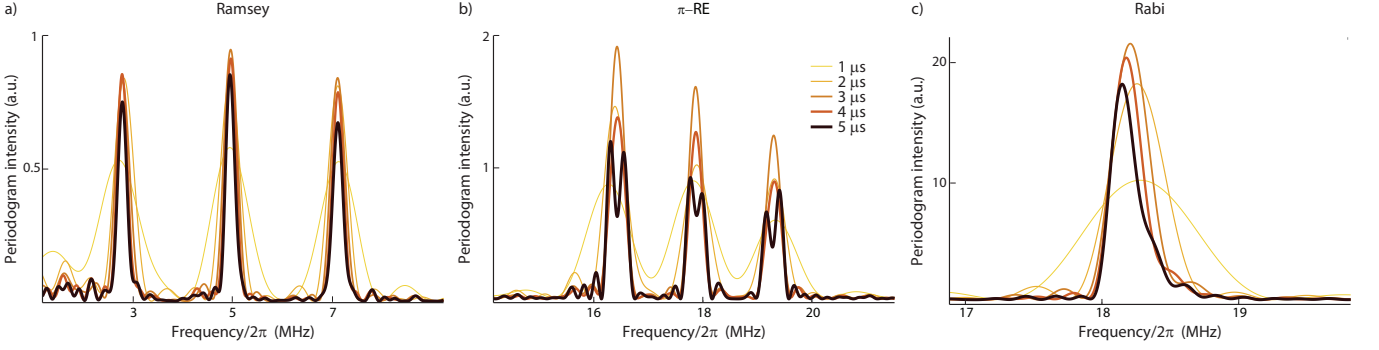}
\caption{\textbf{Experimental periodograms for increasing interrogation times up to $5\mu$s.} a) Ramsey ($5$MHz detuned from the presumed resonance), b) $\pi$-RE and c) Rabi sequences. There is a trade-off between signal intensity and sensitivity to the detunings $\{b, A \pm b\}$ in the signal.} 
\label{fig:SOMcompPer}
\end{figure}

We estimate the statistical significance level $p$ of individual peaks~\cite{Wei06}. Letting $I_m$ be the intensity of $m$-th largest ordinate among the total $M$ in the periodogram, and calculating 
\begin{equation}
T_m = \frac{I_m}{\sum_{k} I_k - \sum_{l=1}^{m}I_l} \ ,
\end{equation}  
the statistical significance of the $m$-th peak $p_m$ is approximated by 
\begin{equation}
p_m \approx (M - (m-1))(1 - T_m)^{M - m} \ .
\end{equation}
To determine $\delta f $, we first estimate the $S/N$ for each periodogram peak by dividing the peak area by the noise floor below the line of $p = 0.01$. 

The periodogram, if plotted over the full spectrum as in Supplementary Fig.~\ref{fig:SOMPeriodoFull}, exhibits very high peaks corresponding to the even harmonics of $\Omega$ which are present in the signal, but which are not taken into account by the first order of average Hamiltonian theory. 
In the inset, the peak structure originated from the detunings of interest is plotted for times much longer than the dephasing time $T'_{RE}$. 

\begin{figure}
\centering
\includegraphics[width=0.5\textwidth]{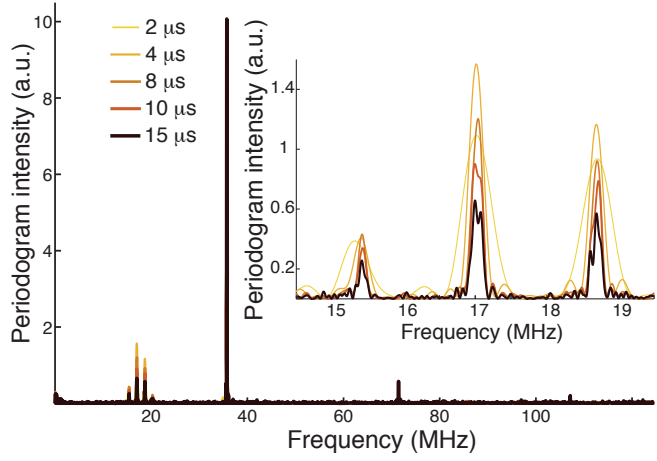}
\caption{\textbf{Periodogram over the full spectrum.} Frequencies corresponding to the even harmonics of $\Omega$ which are present in the signal, but which are not contemplated by the first order of average Hamiltonian theory, can be clearly identified. In the inset, the signal peaks arising from the frequencies of interest are shown for interrogation times much longer than the dephasing time $T'_{RE}$.}
\label{fig:SOMPeriodoFull}
\end{figure}

\section{\label{Sens}Experimental sensitivity}

For a fixed interrogation time $t$, and scanning the detuning from resonance $\delta\omega$, we expect to observe the signal
\begin{equation}
\mathcal{S}(\delta\omega) \propto \cos\left(\frac{2\delta\omega t}{\theta}\sin\left(\frac{\theta}{2}\right)\right) \equiv \cos\left(\delta\omega\tau\right),
\end{equation}
with $\tau=\frac{2t}{\theta}\sin\left(\frac{\theta}{2}\right)$.

In every experimental run, reference curves are acquired along with the signal $\mathcal{S}$; they are noted $\mathcal{R}_0$ for the $|0\rangle$ state as obtained after laser polarization, and $\mathcal{R}_1$ for the $|1\rangle$ state as calibrated by adiabatic inversion. The signal is then normalized as
\begin{equation}
\overline{\mathcal{S}} = \frac{\mathcal{S} - \mathcal{R}_1}{\mathcal{R}_0 - \mathcal{R}_1} \ ;
\end{equation}
The standard deviation of the normalized signal is readily obtained
\begin{equation}
\Delta\overline{\mathcal{S}} = \sqrt{(\Delta \mathcal{R}_0)^2 \left|\frac{\mathcal{S} - \mathcal{R}_1}{(\mathcal{R}_0 - \mathcal{R}_1)^2}\right| ^2 + (\Delta \mathcal{R}_1)^2\left|\frac{\mathcal{S}-\mathcal{R}_1}{(\mathcal{R}_0-\mathcal{R}_1)^2}-\frac{1}{\mathcal{R}_0-\mathcal{R}_1} \right|^2 +  (\Delta \mathcal{S})^2\left|\frac{1}{\mathcal{R}_0-\mathcal{R}_1}\right|^2} \ . 
\end{equation}

The sensitivity is calculated  for the whole signal
\begin{equation}
\eta(\delta\omega) =  \frac{1}{\gamma_e}\frac{\Delta \overline{\mathcal{S}}}{|\frac{\partial \overline{\mathcal{S}}}{\partial \delta\omega}|}\sqrt{N t} \ ;
\end{equation}
for each fixed interrogation time $t$, we single out the minimum sensitivity $\eta(\delta\omega)$ within one period  
of the fitted oscillation period $\tau$, depicted in Supplementary Fig.~\ref{fig:SOMwosc_Ramsey}.a.  

\begin{figure}
\centering
\includegraphics[width=0.75\textwidth]{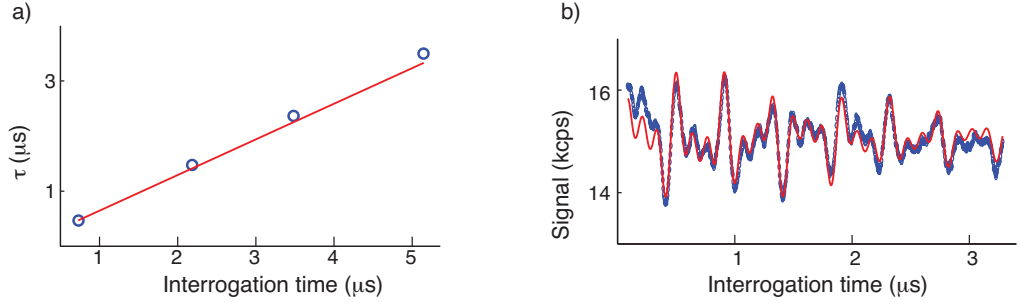}
\caption{\textbf{Fitted parameters used to determine the experimental sensitivity.} a) The fitted periods $\tau$ (blue dots) are linear in $t$ and agree well with the theoretically expected time $\frac{2t\sin(\theta/2)}{\theta}$ (red line). b) The Ramsey signal (blue circles) is fitted to $\mathcal{S}_{Ram} = k_1 - k_2\left[\cos(\delta\omega t) + \cos((A+\delta\omega) t) + \cos((A-\delta\omega) t)\right]e^{-(t/T_2^{\star})^2}$ (red), with fitting parameters $\{k_1,k_2,\delta\omega,A,T_2^{\star}\}$. In particular, $T_2^{\star} \sim 2.19 \pm 0.15\mu$s.}
\label{fig:SOMwosc_Ramsey}
\end{figure}

The standard deviation for the sensitivity measurements is obtained by 
\begin{equation}
\Delta\eta =  \frac{1}{\gamma_e}\left|\frac{\partial \eta}{\partial \overline{\mathcal{S}}}\right|  \Delta\overline{\mathcal{S}}\sqrt{N t} =  \frac{1}{\gamma_e}\left| \frac{1 - 2\overline{\mathcal{S}}}{2\sqrt{\overline{\mathcal{S}}(1-\overline{\mathcal{S}})  }}    \frac{1}{\frac{\partial \overline{\mathcal{S}}}{\partial \delta\omega}}     \right| \Delta\overline{\mathcal{S}}\sqrt{N t}  \ . 
\end{equation}

Additionally, to every point in the plot there corresponds a  factor $C$ taking into account imperfect state detection~\cite{Taylor08,Meriles10}. 
While the theoretical signal $\mathcal{S}$ represents the population in the $\ku$ state, measured from the observable $M \equiv |0\rangle\langle0|$, the experimental signal records photons emitted by 
both $|0\rangle$ and $|1\rangle$ states, so that the measurement operator  is best experimentally described by $M' \equiv n_0|0\rangle\langle0| + n_1|1\rangle\langle1|$. Here, $\{n_0,n_1\}$ are Poisson-distributed variables that indicate the number of collected photons; if perfect state discrimination were possible, $n_0 \rightarrow \infty$ and $n_1 \rightarrow 0$. Including this effect, after $n$ full echo cycles, the signal is modified to
\begin{equation}
{\mathcal{S}'}(n) \approx \frac{1}{4}\left[\left(3n_0 + n_1 + (n_0 - n_1)\cos\theta\right) + \left(n_0 - n_1 - (n_0-n_1)\cos\theta\right)                    \cos\left(\frac{4\delta\omega n}{\Omega}\sin\left(\frac{\theta}{2}\right)\right) \right] \ .
\end{equation}

We calculate the sensitivity for $\overline{\mathcal{S}'}$ in the best-case scenario of minimum sensitivity given by the accumulated phase 
$\left(\frac{4\delta\omega n}{\Omega}\sin\left(\frac{\theta}{2}\right)\right) = \frac{\pi}{2}$, and note the existence of a factor $C$, with respect to the ideal sensitivity, $\eta_{M'} = \eta_M/C$:
\begin{equation}
C^{-1}=\sqrt{1 + \frac{1}{2}+\frac{(-11n_0+5n_1)}{2(n_0-n_1)^2}+\frac{\cos\theta}{2}\left(1-\frac{(n_0+n_1)}{(n_0-n_1)^2}\right)+\frac{8n_0}{(n_0-n_1)^2\sin^2 (\theta/2)}}.
\label{eq:C}
\end{equation}

We use for $n_0$ ($n_1$) the mean photon number for the $|0\rangle$ ($|1\rangle$) reference curve during each acquisition for different $t$. On average, $\overline{n_0} \sim 0.0022 \pm 0.0003$ and $\overline{n_1} \sim 0.0015 \pm 0.0002$.

We also consider the fact that the signal $\overline{\mathcal{S}'}$ has contributions from three detunings $\{\delta\omega, A \pm \delta\omega\}$, where $A$ is the hyperfine coupling between the NV center and spin-1 $^{14}$N nucleus; taking such detunings into account is, incidentally, fundamental for the choice of interrogation times: given the modulation imposed by the multiple frequencies in the signal, full echo times yielding a high signal amplitude are preferred. A different strategy would be to polarize the nuclear spin~\cite{Neumann10}.
In order to compare the ideal sensitivity with the experimental one, in our experiments we need to introduce a further correction factor $C_A$, since the accumulated phase is only equal to the optimal $\frac{\pi}{2}$ for the experimental realizations with $m_I = 0$. We expect the sensitivity to become larger as $\eta_{A} = \eta_{M'}/C_A = \eta_M/(C \times C_A)$, with

\begin{equation}
C_A^{-1} = \frac{3}{\left|1+2\cos\left(\frac{2At\sin(\theta/2)}{\theta}\right)\right|}.
\end{equation}

In order to estimate $C \times C_A$, we use the fitted value for $A$ at each point ($\overline{A} \sim 2 \pi \times (2.21 \pm 0.07)$MHz), the time $t$ corresponding to the number of cycles at which the experimental point was taken, and a corrected $\theta \sim 0.984 \pi$ that takes into account the real angle, given the experimental Rabi frequency, imposed by the duration of the echo half cycle, which can be controlled only up to the inverse of the AWG sample rate. A mean total correction factor of $\overline{C \times C_A} \sim (5.9\pm1.4) \times 10^{-3}$ is obtained for the set of points. The mean sensitivity curve (solid line) is expressed as the theoretically expected sensitivity in the absence of noise $\eta_M$, divided by $\overline{C \times C_A}$. Similarly, the lower (higher) bounds for the sensitivity are estimated by dividing the theoretical sensitivity by the maximum (minimum) $C \times C_A$ value in the set of points.

The effect of decoherence is included in the plot using a fit for $T_2^{\star} \sim 2.19 \pm 0.15\mu$s from the Ramsey experiment shown in Supplementary Fig.~\ref{fig:SOMwosc_Ramsey}.b. Assuming static Gaussian noise, we let $\eta_{A} \rightarrow \eta_{A} e^{(t/T'_{RE})^2}$, with $T'_{RE} = \frac{T_2^{\star}\theta}{2\sin(\theta/2)}$. 

\section{\label{Rabi}Rabi-beat magnetometry}

Rabi-beat magnetometry using a single solid-state qubit was recently demonstrated~\cite{Fedder11}. The scheme presupposes the existence of an absolute frequency standard against which one wishes to resolve a nearby frequency. For magnetometry purposes then,
\begin{equation}
\mathcal{S} = \frac{1}{2}(\mathcal{S}_{Rabi}(\delta\omega) - \mathcal{S}_{Rabi}(0)) \ ,
\end{equation}
where $\delta\omega$ denotes a detuning from the frequency standard. The sensitivity reads
\begin{equation}
\eta = \frac{1}{\gamma_e}\lim_{\delta \omega \rightarrow 0} \frac{\Delta \mathcal{S}}{|\frac{\partial \mathcal{S}}{\partial \delta\omega}|} \sqrt{t} \approx  \frac{\sqrt{2\Omega}}{\gamma_e}  \sqrt{\frac{t\Omega}{2 - 2\cos(t\Omega) - t\Omega \sin(t\Omega)}} \ ;
\end{equation}
$\eta$ is close to minima at $t \approx (2k + \frac{3}{2})\frac{\pi}{\Omega}$, yielding
\begin{equation}
\eta_{min} \approx \frac{\sqrt{2\Omega}}{\gamma_e}\Big/\sqrt{1 + \frac{2}{t\Omega}} \ ,
\end{equation}
which tends to $\sqrt{2\Omega}/\gamma_e$ for increasingly large interrogation times.

\section{\label{Znoise}Evolution under bath noise}


In the presence of Gaussian static noise  in the $z$-direction with variance $\sigma^2$, the RE signal decays as 
\begin{equation}
\langle \mathcal{S}_{RE} \rangle=\frac{1}{2}\left[1 + \cos^2\left(\frac{\theta}{2}\right) + \sin^2\left(\frac{\theta}{2}\right)\cos\left(\frac{2\delta\omega t}{\theta}\sin\left(\frac{\theta}{2}\right)\right)e^{(t/T'_{RE})^2}\right],
\end{equation}
where we define the dephasing time
\begin{equation}
T'_{RE} = \frac{\theta}{\sigma\sqrt{2}|\sin(\theta/2)|}.
\end{equation}
Similarly, one obtains for the Ramsey signal
\begin{equation}
\langle \mathcal{S}_{Ram} \rangle =\half\left(1+\cos(\delta\omega t) e^{-(t/T'_{Ram})^2}\right) \ , \quad \text{with} \quad T'_{Ram} = T_2^{\star} = \frac{\sqrt{2}}{\sigma} \ .
\end{equation}
Note that $T'_{RE} > T'_{Ram}$ always; nevertheless, at the optimum interrogation time calculated for both sequences as $\frac{T'}{2}$, 
\begin{equation}
\frac{\eta_{RE}}{\eta_{Ram}} = \sqrt{\frac{\theta}{2\sin(\theta/2)^3}} > 1 \ ;
\end{equation}
the sensitivity ratio above has a minimum $\eta_{RE}/\eta_{Ram}\sim 1.20$ for $\theta \sim \frac{3\pi}{4}$, which is the angle that yields the highest sensitivity for the RE sequence.

We now turn our attention to the evolution of the Rabi signal under Gaussian dephasing noise. For $\delta\omega \ll \Omega$, the Rabi signal is approximately
\begin{equation}
\mathcal{S}_{Rabi} = 1 - \frac{\Omega^2}{\Omega^2 + \delta\omega^2}\sin^2\left(\frac{t}{2}\sqrt{\Omega^2 + \delta\omega^2} \right) \approx 1 - \left( 1 - \frac{\delta \omega ^2}{\Omega^2}  \right) \sin^2  \left( \frac{t}{2} \left(\Omega + \frac{\delta\omega^2}{\Omega}\right)   \right)  \ ;    
\end{equation}
calculating the expected value $\langle \mathcal{S}_{Rabi} \rangle$ under the noise distribution yields
\begin{equation}
\langle \mathcal{S}_{Rabi} \rangle = \frac{1}{2} \left[1 +  \frac{\cos(t\Omega + \arctan(t \sigma^2/\Omega)/2)}{\left(1+ \frac{t^2\sigma^4}{\Omega^2} \right)^{\frac{1}{4}}}   + \frac{\sigma^2}{\Omega^2}\left( 1 - \frac{\cos(t\Omega + 3\arctan(t \sigma^2/\Omega)/2)}{\left(1+ \frac{t^2\sigma^4}{\Omega^2} \right)^{\frac{3}{4}}}\right)\right] \ .
\end{equation}


In the presence of stochastic (Ornstein-Uhlenbeck)  noise with zero mean and autocorrelation function $\sigma^2 e^{-\frac{t}{\tau_c}}$, a Ramsey signal decays as~\cite{Kubo62}
\begin{equation}
\ave{ \mathcal{S}_{Ram}}=\frac{1}{2}\left(1 + e^{-\zeta'(t)}\right) \ , \quad  \text{with}\quad   \zeta'(t) =\sigma^2 \tau_c^2 (t/\tau_c + e^{-\frac{t}{\tau_c}} - 1) \ .
\end{equation}

Numerical simulations valid for $\tau_c \sigma \lesssim \theta/2$ and $\tau_c \gtrsim \theta/(2\Omega)$ indicate that the RE signal decays as 
\begin{equation}
\langle\mathcal{S}_{RE}\rangle = \frac{1}{2}\left[1 + \cos^2\left(\frac{\theta}{2}\right) + \sin^2\left(\frac{\theta}{2}\right) e^{-\zeta(t)}\right] \ ,
\end{equation}
with
\begin{equation}
\zeta(t) = \zeta'(t)\frac{4\sin^2(\theta/2)}{\theta^2} \ .
\end{equation}
Note the additional factor $\frac{4\sin^2(\theta/2)}{\theta^2} = \left(\frac{T'_{Ram}}{T'_{RE}}\right)^2 < 1$. 

Simulations that compare different $\theta$-RE for $\theta = \{3\pi/4, \pi, 5\pi\}$ and Ramsey signals in the presence of stochastic noise are depicted in Supplementary Fig.~\ref{fig:SOMcompNOISE}.a; it is clear that RE sequences are more resilient to bath noise with correlation times shorter than the echo period.

Previous calculations~\cite{Dobrovitski09} indicate that, for slow baths $\frac{1}{\tau_c} \ll \frac{\sigma^2}{\Omega}$, the Rabi signal follows the static noise behaviour for short times, and decays $\propto e^{-\frac{\sigma t}{2\sqrt{\tau_c \Omega}}}$ for long times. Fast baths $\frac{1}{\tau_c} \gg \frac{\sigma^2}{\Omega}$ induce a decay of the Rabi signal $\propto e^{\frac{4 \Omega^2}{\sigma^4 \tau_c}}$.

\begin{figure}
\centering
\includegraphics[width=\textwidth]{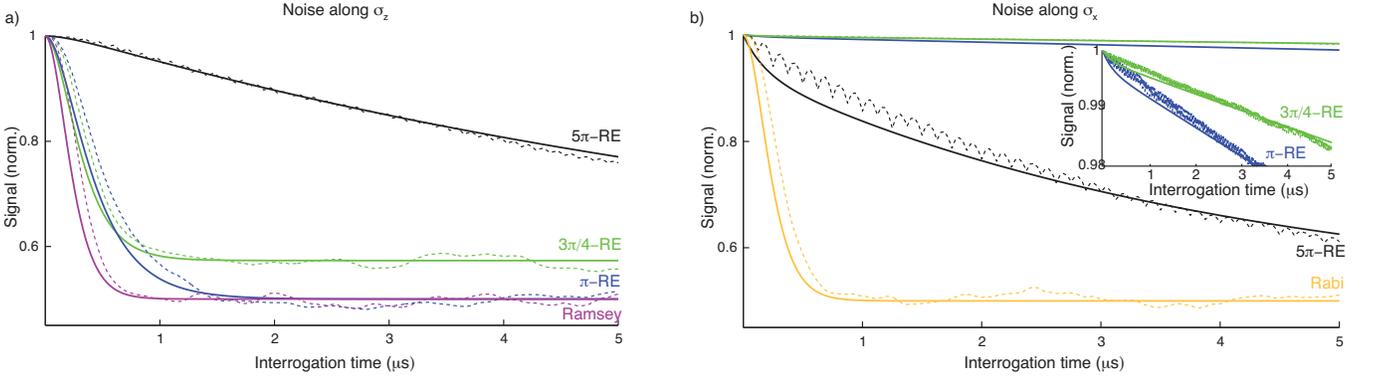}
\caption{\textbf{Signal decay in the presence of stochastic noise}. a) In the presence of stochastic bath noise, different $\theta$-RE with $\theta = \{3\pi/4, \pi, 5\pi\}$ (green, blue, black) are compared against a Ramsey sequence (purple); numerical simulations (dashed lines) agree with the formulas presented in the text (solid lines). The Ramsey sequence is the least resilient to bath noise, whereas one can adjust the dephasing of the RE by the choice of $\theta$; RE sequences are more resilient to bath noise with correlation times shorter than the echo period. The used numerical parameters are: $\Omega = 2\pi\!\times\!20$MHz, $\delta\omega =  2\pi\!\times\!2$MHz, $\tau_c = 200$ns, $\sigma = 0.05 \Omega$. b) In the presence of stochastic noise in the excitation field, the situation is inverted: RE sequences refocus microwave noise with correlation times longer than the echo period; the decay of the Rabi sequence (orange) is plotted for comparison. Used parameters are the same as in a), except for $\delta\omega = 0$.}
\label{fig:SOMcompNOISE}
\end{figure}

\section{\label{Xnoise}Evolution under excitation field noise}


In the presence of a constant error in the Rabi frequency such that $\Omega \rightarrow (1 + \epsilon)\Omega$, the infidelity $(1 - \text{Tr}[U(\epsilon)U(0)]/2) \equiv (1 - F)$ of the pulse sequence is given to second order in the detuning from resonance $\delta \omega$ and in $\epsilon$ by
\begin{equation}
(1 - F)_{RE} \approx \frac{\epsilon^2 t^2 \delta\omega^2}{8}\frac{(2 + \theta^2 - 2\cos\theta - 2\theta\sin\theta)}{\theta^2}
\end{equation}
for RE and 
\begin{equation}
(1 - F)_{Rabi} \approx \frac{\epsilon^2 t^2 \Omega^2}{8} - \frac{\epsilon^2\delta\omega^2(-2 + t^2\Omega^2 + 2\cos(t\Omega))}{8\Omega^2}
\end{equation}
for Rabi-beat magnetometry.

Similarly, an error in the Rabi frequency will yield a flip-angle error in the Ramsey sequence, resulting in the infidelity
\begin{equation}
(1-F)_{Ram}\approx \frac{\epsilon^2 \pi^2}{8} - \frac{\epsilon^2\delta\omega^2(-16 + 4\pi^2 + \pi t \Omega (8 + \pi t \Omega))}{32\Omega^2} \ .
\end{equation}


In the presence of stochastic noise in the excitation field with zero mean and autocorrelation function $\sigma^2e^{-\frac{t}{\tau_c}}$, the resonant cases for RE, Rabi have simple analytical solutions.   

A cumulant expansion technique applied to periodic Hamiltonians~\cite{Kubo62,Cappellaro06}  yields for the envelope of a resonant RE sequence
\begin{equation}
\langle\mathcal{S}_{RE}\rangle = \frac{1}{2}\left(1 + e^{-\zeta(n)}\right) \ ,
\end{equation}
with
\begin{equation}
\zeta(n) = \tau_c^2\sigma^2 \left[ \frac{2n\theta}{\sigma\tau_c} + 2n(e^{-\frac{\theta}{\Omega\tau_c}} - 1) -\tanh^2\left(\frac{1}{2}\frac{\theta}{\Omega\tau_c}    \right) \left(2n(e^{-\frac{\theta}{\Omega\tau_c}} +1) + e^{-\frac{2n\theta}{\tau_c\sigma}} -1\right)\right] \ .
\end{equation}
We note that this decay is equivalent to the decay under pure dephasing for a PDD sequence~\cite{Khodjasteh05}. 

In Supplementary Fig.~\ref{fig:SOMcompNOISE}.b, we simulate the signal for different $\theta$-RE and Rabi sequences if noise in the excitation field is present. Contrarily to RE decay in the presence of bath noise, and as shown experimentally in the main text, RE sequences can refocus excitation noise with correlation times longer than the echo period.

We  note  that the Rabi signal decay for noise along $\sigma_x$ should be comparable to Ramsey signal decay in the presence of stochastic noise along $\sigma_z$. We thus have the decay
\begin{equation}
\ave{ \mathcal{S}_{Rabi}}=\frac{1}{2}\left(1 + e^{-\zeta'(t)}\right) \ , \quad  \text{with}\quad   \zeta'(t) =\sigma^2 \tau_c^2 (t/\tau_c + e^{-\frac{t}{\tau_c}} - 1) \ .
\end{equation}
The advantage of the RE sequence over the Rabi is thus the same advantage that dynamical decoupling sequences can offer.
\section{Repeated readouts}
 The NV spin state can be  read under non-resonant illumination at room temperature  using the fact that the $m_s=\pm1$ excited states can decay  into  metastable states, which live for $\sim$300~ns, while direct optical decay  happens in about 12~ns.  Thus, a NV in the $m_s=0$ state will emit, and absorb, approximately 15 photons, compared to only a few for a NV in the $m_s=\pm1$  states, yielding  state discrimination by fluorescence intensity.
Unfortunately the metastable state decays primarily via spin-non conserving processes into the $m_s=0$ state thereby re-orienting the spin.  This is good for spin polarization, but erases the spin memory and reduces measurement contrast.  The detection efficiency $C$ of the NV center spin state is thus given by Eq.~\ref{eq:C}, which for $\theta=k\pi$ reduces to $C=\left(1+\frac{3(n_0+n_1)}{(n_0-n_1)^2}\right)^{-1/2}$, where $n_{1,0}$ is the number of photons collected if the NV spin is in the $m_s=\{0,1\}$ state, respectively. 

In the repeated readout scheme~\cite{Jiang09,Neumann10b}, the state of the nuclear spin is repetitively mapped onto the electronic spin, which is then read out under laser illumination. The measurement projects the nuclear spin state into a mixed state, but the information about its population difference is preserved, under the assumption that the measurement is a good quantum non-demolition measurement. We can include the effect of these repeated readout by defining a new detection efficiency, $C_{N_r}=\left(1+\frac{1}{N_r}\frac{3(n_0+n_1)}{(n_0-n_1)^2}\right)^{-1/2}$, which shows an improvement $\propto\sqrt{N_r}$, where $N_r$ is the number of measurements. The sensitivity needs of course to be further modified to take into account the increased measurement time. Provided the time needed for one measurement step is smaller than the interrogation time, it becomes advantageous to use repeated readouts.  As RE increases the interrogation time, it can achieve better sensitivity than Ramsey magnetometry by using the repeated readout scheme (which is instead not advantageous for a simple Ramsey scheme), as depicted in Supplementary Fig.~\ref{fig:RErepeatedreadouts}. 
\begin{figure}
\centering
\includegraphics[width=0.5\textwidth]{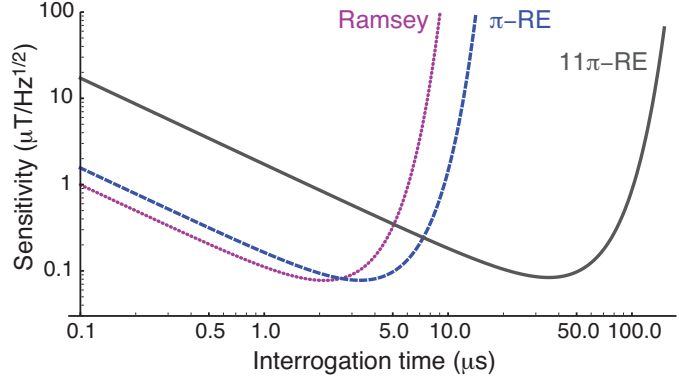}
\caption{\textbf{Sensitivity with repeated readouts.} We compare the achievable sensitivity for Ramsey (purple, dotted) and Rotary echo ($\pi$-RE, blue, dashed; $11\pi$-RE, gray) sequences, when using the repeated readout scheme, with $N_r=100$ and the time per each readout $t_r=1.5\mu$s. In the presence of dephasing with $T_2^\star=3\mu$s, the longer angle RE achieves good sensitivity for much longer interrogation times.} 
\label{fig:RErepeatedreadouts}
\end{figure}

\section{\label{Ca}Calcium signaling domains}
Although virtually all neuronal reactions are regulated by diffusing Ca$^{2+}$ ions between membrane channel sources and cytoplasm target receptors, triggering specificity is ensured by the fact that such diffusion events are localized in time and space. The size of the signaling domain, understood as roughly the distance between channel and receptor ($50$nm to $0.5\mu$m), determines the diffusion timescale ($\mu$s to ms) and strength, the latter measured by Ca$^{2+}$ concentration ($100$ to $1\mu$M)~\cite{Augustine03,Fakler08}.

Let a flux with duration $t$ and mean travelled distance $d$ between membrane channel and cytoplasm receptor. The magnetic field at a distance $r$ from a transient Ca$^{+2}$ flux composed of $l$ ions is estimated as 
\begin{equation}
B (T) = \frac{\mu_0}{4\pi}\frac{2 l e d}{t r^2} \ ,
\end{equation}
with $e$ the electron charge, and with the magnetic permeability of the cell approximated by $\mu_0$, the vacuum permeability. Therefore, the minimum required sensitivity to sense the afore-described calcium flux is
\begin{equation}
\eta \left(\frac{T}{\sqrt{\textrm{Hz}}}\right) = \sqrt{2\pi}\frac{\mu_0}{4\pi}\frac{2 l e d}{\sqrt{t} r^2}\sqrt{N} \ .
\end{equation}


\begin{thebibliography}{33}%
\makeatletter
\providecommand \@ifxundefined [1]{%
 \@ifx{#1\undefined}
}%
\providecommand \@ifnum [1]{%
 \ifnum #1\expandafter \@firstoftwo
 \else \expandafter \@secondoftwo
 \fi
}%
\providecommand \@ifx [1]{%
 \ifx #1\expandafter \@firstoftwo
 \else \expandafter \@secondoftwo
 \fi
}%
\providecommand \natexlab [1]{#1}%
\providecommand \enquote  [1]{``#1''}%
\providecommand \bibnamefont  [1]{#1}%
\providecommand \bibfnamefont [1]{#1}%
\providecommand \citenamefont [1]{#1}%
\providecommand \href@noop [0]{\@secondoftwo}%
\providecommand \href [0]{\begingroup \@sanitize@url \@href}%
\providecommand \@href[1]{\@@startlink{#1}\@@href}%
\providecommand \@@href[1]{\endgroup#1\@@endlink}%
\providecommand \@sanitize@url [0]{\catcode `\\12\catcode `\$12\catcode
  `\&12\catcode `\#12\catcode `\^12\catcode `\_12\catcode `\%12\relax}%
\providecommand \@@startlink[1]{}%
\providecommand \@@endlink[0]{}%
\providecommand \url  [0]{\begingroup\@sanitize@url \@url }%
\providecommand \@url [1]{\endgroup\@href {#1}{\urlprefix }}%
\providecommand \urlprefix  [0]{URL }%
\providecommand \Eprint [0]{\href }%
\providecommand \doibase [0]{http://dx.doi.org/}%
\providecommand \selectlanguage [0]{\@gobble}%
\providecommand \bibinfo  [0]{\@secondoftwo}%
\providecommand \bibfield  [0]{\@secondoftwo}%
\providecommand \translation [1]{[#1]}%
\providecommand \BibitemOpen [0]{}%
\providecommand \bibitemStop [0]{}%
\providecommand \bibitemNoStop [0]{.\EOS\space}%
\providecommand \EOS [0]{\spacefactor3000\relax}%
\providecommand \BibitemShut  [1]{\csname bibitem#1\endcsname}%
\let\auto@bib@innerbib\@empty
\bibitem [{\citenamefont {Taylor}\ \emph {et~al.}(2008)\citenamefont {Taylor},
  \citenamefont {Cappellaro}, \citenamefont {Childress}, \citenamefont {Jiang},
  \citenamefont {Budker}, \citenamefont {Hemmer}, \citenamefont {Yacoby},
  \citenamefont {Walsworth},\ and\ \citenamefont {Lukin}}]{Taylor08}%
  \BibitemOpen
  \bibfield  {author} {\bibinfo {author} {\bibfnamefont {J.~M.}\ \bibnamefont
  {Taylor}}, \bibinfo {author} {\bibfnamefont {P.}~\bibnamefont {Cappellaro}},
  \bibinfo {author} {\bibfnamefont {L.}~\bibnamefont {Childress}}, \bibinfo
  {author} {\bibfnamefont {L.}~\bibnamefont {Jiang}}, \bibinfo {author}
  {\bibfnamefont {D.}~\bibnamefont {Budker}}, \bibinfo {author} {\bibfnamefont
  {P.~R.}\ \bibnamefont {Hemmer}}, \bibinfo {author} {\bibfnamefont
  {A.}~\bibnamefont {Yacoby}}, \bibinfo {author} {\bibfnamefont
  {R.}~\bibnamefont {Walsworth}}, \ and\ \bibinfo {author} {\bibfnamefont
  {M.~D.}\ \bibnamefont {Lukin}},\ }\href {\doibase 10.1038/nphys1075}
  {\bibfield  {journal} {\bibinfo  {journal} {Nature Phys.}\ }\textbf {\bibinfo
  {volume} {4}},\ \bibinfo {pages} {810} (\bibinfo {year} {2008})}\BibitemShut
  {NoStop}%
\bibitem [{\citenamefont {Levitt}(1986)}]{Levitt86}%
  \BibitemOpen
  \bibfield  {author} {\bibinfo {author} {\bibfnamefont {M.~H.}\ \bibnamefont
  {Levitt}},\ }\href {\doibase 10.1016/0079-6565(86)80005-X} {\bibfield
  {journal} {\bibinfo  {journal} {Prog. Nucl. Mag. Res. Spect.}\ }\textbf
  {\bibinfo {volume} {18}},\ \bibinfo {pages} {61} (\bibinfo {year}
  {1986})}\BibitemShut {NoStop}%
\bibitem [{\citenamefont {Solomon}(1958)}]{Solomon57}%
  \BibitemOpen
  \bibfield  {author} {\bibinfo {author} {\bibfnamefont {I.}~\bibnamefont
  {Solomon}},\ }\href {\doibase 10.1103/PhysRev.110.61} {\bibfield  {journal}
  {\bibinfo  {journal} {Phys. Rev.}\ }\textbf {\bibinfo {volume} {110}},\
  \bibinfo {pages} {61} (\bibinfo {year} {1958})}\BibitemShut {NoStop}%
\bibitem [{\citenamefont {Maze}\ \emph {et~al.}(2008)\citenamefont {Maze},
  \citenamefont {Stanwix}, \citenamefont {Hodges}, \citenamefont {Hong},
  \citenamefont {Taylor}, \citenamefont {Cappellaro}, \citenamefont {Jiang},
  \citenamefont {Zibrov}, \citenamefont {Yacoby}, \citenamefont {Walsworth},\
  and\ \citenamefont {Lukin}}]{Maze08}%
  \BibitemOpen
  \bibfield  {author} {\bibinfo {author} {\bibfnamefont {J.~R.}\ \bibnamefont
  {Maze}}, \bibinfo {author} {\bibfnamefont {P.~L.}\ \bibnamefont {Stanwix}},
  \bibinfo {author} {\bibfnamefont {J.~S.}\ \bibnamefont {Hodges}}, \bibinfo
  {author} {\bibfnamefont {S.}~\bibnamefont {Hong}}, \bibinfo {author}
  {\bibfnamefont {J.~M.}\ \bibnamefont {Taylor}}, \bibinfo {author}
  {\bibfnamefont {P.}~\bibnamefont {Cappellaro}}, \bibinfo {author}
  {\bibfnamefont {L.}~\bibnamefont {Jiang}}, \bibinfo {author} {\bibfnamefont
  {A.}~\bibnamefont {Zibrov}}, \bibinfo {author} {\bibfnamefont
  {A.}~\bibnamefont {Yacoby}}, \bibinfo {author} {\bibfnamefont
  {R.}~\bibnamefont {Walsworth}}, \ and\ \bibinfo {author} {\bibfnamefont
  {M.~D.}\ \bibnamefont {Lukin}},\ }\href {\doibase 10.1038/nature07279}
  {\bibfield  {journal} {\bibinfo  {journal} {Nature}\ }\textbf {\bibinfo
  {volume} {455}},\ \bibinfo {pages} {644} (\bibinfo {year}
  {2008})}\BibitemShut {NoStop}%
\bibitem [{\citenamefont {Balasubramanian}\ \emph {et~al.}(2008)\citenamefont
  {Balasubramanian}, \citenamefont {Chan}, \citenamefont {Kolesov},
  \citenamefont {Al-Hmoud}, \citenamefont {Shin}, \citenamefont {Kim},
  \citenamefont {Wojcik}, \citenamefont {Hemmer}, \citenamefont
  {Kr$\ddot{u}$ger}, \citenamefont {Jelezko},\ and\ \citenamefont
  {Wrachtrup}}]{Balasubramanian08}%
  \BibitemOpen
  \bibfield  {author} {\bibinfo {author} {\bibfnamefont {G.}~\bibnamefont
  {Balasubramanian}}, \bibinfo {author} {\bibfnamefont {I.-Y.}\ \bibnamefont
  {Chan}}, \bibinfo {author} {\bibfnamefont {R.}~\bibnamefont {Kolesov}},
  \bibinfo {author} {\bibfnamefont {M.}~\bibnamefont {Al-Hmoud}}, \bibinfo
  {author} {\bibfnamefont {C.}~\bibnamefont {Shin}}, \bibinfo {author}
  {\bibfnamefont {C.}~\bibnamefont {Kim}}, \bibinfo {author} {\bibfnamefont
  {A.}~\bibnamefont {Wojcik}}, \bibinfo {author} {\bibfnamefont {P.~R.}\
  \bibnamefont {Hemmer}}, \bibinfo {author} {\bibfnamefont {A.}~\bibnamefont
  {Kr$\ddot{u}$ger}}, \bibinfo {author} {\bibfnamefont {F.}~\bibnamefont
  {Jelezko}}, \ and\ \bibinfo {author} {\bibfnamefont {J.}~\bibnamefont
  {Wrachtrup}},\ }\href {\doibase 10.1038/nature07278} {\bibfield  {journal}
  {\bibinfo  {journal} {Nature}\ }\textbf {\bibinfo {volume} {445}},\ \bibinfo
  {pages} {648} (\bibinfo {year} {2008})}\BibitemShut {NoStop}%
\bibitem [{\citenamefont {Fedder}\ \emph {et~al.}(2011)\citenamefont {Fedder},
  \citenamefont {Dolde}, \citenamefont {Rempp}, \citenamefont {Wolf},
  \citenamefont {Hemmer}, \citenamefont {Jelezko},\ and\ \citenamefont
  {Wrachtrup}}]{Fedder11}%
  \BibitemOpen
  \bibfield  {author} {\bibinfo {author} {\bibfnamefont {H.}~\bibnamefont
  {Fedder}}, \bibinfo {author} {\bibfnamefont {F.}~\bibnamefont {Dolde}},
  \bibinfo {author} {\bibfnamefont {F.}~\bibnamefont {Rempp}}, \bibinfo
  {author} {\bibfnamefont {T.}~\bibnamefont {Wolf}}, \bibinfo {author}
  {\bibfnamefont {P.}~\bibnamefont {Hemmer}}, \bibinfo {author} {\bibfnamefont
  {F.}~\bibnamefont {Jelezko}}, \ and\ \bibinfo {author} {\bibfnamefont
  {J.}~\bibnamefont {Wrachtrup}},\ }\href {\doibase 10.1007/s00340-011-4408-4}
  {\bibfield  {journal} {\bibinfo  {journal} {Applied Physics B: Lasers and
  Optics}\ }\textbf {\bibinfo {volume} {102}},\ \bibinfo {pages} {497}
  (\bibinfo {year} {2011})}\BibitemShut {NoStop}%
\bibitem [{\citenamefont {{Jarmola}}\ \emph {et~al.}(2011)\citenamefont
  {{Jarmola}}, \citenamefont {{Acosta}}, \citenamefont {{Jensen}},
  \citenamefont {{Chemerisov}},\ and\ \citenamefont {{Budker}}}]{Jarmola12x}%
  \BibitemOpen
  \bibfield  {author} {\bibinfo {author} {\bibfnamefont {A.}~\bibnamefont
  {{Jarmola}}}, \bibinfo {author} {\bibfnamefont {V.~M.}\ \bibnamefont
  {{Acosta}}}, \bibinfo {author} {\bibfnamefont {K.}~\bibnamefont {{Jensen}}},
  \bibinfo {author} {\bibfnamefont {S.}~\bibnamefont {{Chemerisov}}}, \ and\
  \bibinfo {author} {\bibfnamefont {D.}~\bibnamefont {{Budker}}},\ }\href
  {http://arxiv.org/abs/1112.5936} {\bibfield  {journal} {\bibinfo  {journal}
  {ArXiv e-prints}\ } (\bibinfo {year} {2011})},\ \Eprint
  {http://arxiv.org/abs/1112.5936} {arXiv:1112.5936} \BibitemShut {NoStop}%
\bibitem [{\citenamefont {Waldherr}\ \emph {et~al.}(2012)\citenamefont
  {Waldherr}, \citenamefont {Beck}, \citenamefont {Neumann}, \citenamefont
  {Said}, \citenamefont {Nitsche}, \citenamefont {Markham}, \citenamefont
  {Twitchen}, \citenamefont {Twamley}, \citenamefont {Jelezko},\ and\
  \citenamefont {Wrachtrup}}]{Waldherr12}%
  \BibitemOpen
  \bibfield  {author} {\bibinfo {author} {\bibfnamefont {G.}~\bibnamefont
  {Waldherr}}, \bibinfo {author} {\bibfnamefont {J.}~\bibnamefont {Beck}},
  \bibinfo {author} {\bibfnamefont {P.}~\bibnamefont {Neumann}}, \bibinfo
  {author} {\bibfnamefont {R.}~\bibnamefont {Said}}, \bibinfo {author}
  {\bibfnamefont {M.}~\bibnamefont {Nitsche}}, \bibinfo {author} {\bibfnamefont
  {M.}~\bibnamefont {Markham}}, \bibinfo {author} {\bibfnamefont {D.~J.}\
  \bibnamefont {Twitchen}}, \bibinfo {author} {\bibfnamefont {J.}~\bibnamefont
  {Twamley}}, \bibinfo {author} {\bibfnamefont {F.}~\bibnamefont {Jelezko}}, \
  and\ \bibinfo {author} {\bibfnamefont {J.}~\bibnamefont {Wrachtrup}},\ }\href
  {\doibase 10.1038/nnano.2011.224} {\bibfield  {journal} {\bibinfo  {journal}
  {Nat Nano}\ }\textbf {\bibinfo {volume} {7}},\ \bibinfo {pages} {105}
  (\bibinfo {year} {2012})}\BibitemShut {NoStop}%
\bibitem [{\citenamefont {Childress}\ \emph {et~al.}(2006)\citenamefont
  {Childress}, \citenamefont {Gurudev~Dutt}, \citenamefont {Taylor},
  \citenamefont {Zibrov}, \citenamefont {Jelezko}, \citenamefont {Wrachtrup},
  \citenamefont {Hemmer},\ and\ \citenamefont {Lukin}}]{Childress06}%
  \BibitemOpen
  \bibfield  {author} {\bibinfo {author} {\bibfnamefont {L.}~\bibnamefont
  {Childress}}, \bibinfo {author} {\bibfnamefont {M.~V.}\ \bibnamefont
  {Gurudev~Dutt}}, \bibinfo {author} {\bibfnamefont {J.~M.}\ \bibnamefont
  {Taylor}}, \bibinfo {author} {\bibfnamefont {A.~S.}\ \bibnamefont {Zibrov}},
  \bibinfo {author} {\bibfnamefont {F.}~\bibnamefont {Jelezko}}, \bibinfo
  {author} {\bibfnamefont {J.}~\bibnamefont {Wrachtrup}}, \bibinfo {author}
  {\bibfnamefont {P.~R.}\ \bibnamefont {Hemmer}}, \ and\ \bibinfo {author}
  {\bibfnamefont {M.~D.}\ \bibnamefont {Lukin}},\ }\href {\doibase
  10.1126/science.1131871} {\bibfield  {journal} {\bibinfo  {journal}
  {Science}\ }\textbf {\bibinfo {volume} {314}},\ \bibinfo {pages} {281}
  (\bibinfo {year} {2006})}\BibitemShut {NoStop}%
\bibitem [{\citenamefont {Cappellaro}\ \emph {et~al.}(2012)\citenamefont
  {Cappellaro}, \citenamefont {Goldstein}, \citenamefont {Hodges},
  \citenamefont {Jiang}, \citenamefont {Maze}, \citenamefont {S\o{}rensen},\
  and\ \citenamefont {Lukin}}]{Cappellaro12b}%
  \BibitemOpen
  \bibfield  {author} {\bibinfo {author} {\bibfnamefont {P.}~\bibnamefont
  {Cappellaro}}, \bibinfo {author} {\bibfnamefont {G.}~\bibnamefont
  {Goldstein}}, \bibinfo {author} {\bibfnamefont {J.~S.}\ \bibnamefont
  {Hodges}}, \bibinfo {author} {\bibfnamefont {L.}~\bibnamefont {Jiang}},
  \bibinfo {author} {\bibfnamefont {J.~R.}\ \bibnamefont {Maze}}, \bibinfo
  {author} {\bibfnamefont {A.~S.}\ \bibnamefont {S\o{}rensen}}, \ and\ \bibinfo
  {author} {\bibfnamefont {M.~D.}\ \bibnamefont {Lukin}},\ }\href {\doibase
  10.1103/PhysRevA.85.032336} {\bibfield  {journal} {\bibinfo  {journal} {Phys.
  Rev. A}\ }\textbf {\bibinfo {volume} {85}},\ \bibinfo {pages} {032336}
  (\bibinfo {year} {2012})}; 
    \bibfield  {author} {\bibinfo {author} {\bibfnamefont {G.}~\bibnamefont
  {Goldstein}}, \bibinfo {author} {\bibfnamefont {P.}~\bibnamefont
  {Cappellaro}}, \bibinfo {author} {\bibfnamefont {J.~R.}\ \bibnamefont
  {{Maze}}}, \bibinfo {author} {\bibfnamefont {J.~S.}\ \bibnamefont {Hodges}},
  \bibinfo {author} {\bibfnamefont {L.}~\bibnamefont {Jiang}}, \bibinfo
  {author} {\bibfnamefont {A.~S.}\ \bibnamefont {{S{\o}rensen}}}, \ and\
  \bibinfo {author} {\bibfnamefont {M.~D.}\ \bibnamefont {Lukin}},\ }\href
  {\doibase 10.1103/PhysRevLett.106.140502} {\bibfield  {journal} {\bibinfo
  {journal} {Phys. Rev. Lett.}\ }\textbf {\bibinfo {volume} {106}},\ \bibinfo
  {pages} {140502} (\bibinfo {year} {2011})}\BibitemShut {NoStop}%
\bibitem [{\citenamefont {McGuinness}\ \emph {et~al.}(2011)\citenamefont
  {McGuinness}, \citenamefont {Yan}, \citenamefont {Stacey}, \citenamefont
  {Simpson}, \citenamefont {Hall}, \citenamefont {Maclaurin}, \citenamefont
  {Prawer}, \citenamefont {Mulvaney}, \citenamefont {Wrachtrup}, \citenamefont
  {Caruso}, \citenamefont {Scholten},\ and\ \citenamefont
  {Hollenberg}}]{McGuinness11}%
  \BibitemOpen
  \bibfield  {author} {\bibinfo {author} {\bibfnamefont {L.~P.}\ \bibnamefont
  {McGuinness}}, \bibinfo {author} {\bibfnamefont {Y.}~\bibnamefont {Yan}},
  \bibinfo {author} {\bibfnamefont {A.}~\bibnamefont {Stacey}}, \bibinfo
  {author} {\bibfnamefont {D.~A.}\ \bibnamefont {Simpson}}, \bibinfo {author}
  {\bibfnamefont {L.~T.}\ \bibnamefont {Hall}}, \bibinfo {author}
  {\bibfnamefont {D.}~\bibnamefont {Maclaurin}}, \bibinfo {author}
  {\bibfnamefont {S.}~\bibnamefont {Prawer}}, \bibinfo {author} {\bibfnamefont
  {P.}~\bibnamefont {Mulvaney}}, \bibinfo {author} {\bibfnamefont
  {J.}~\bibnamefont {Wrachtrup}}, \bibinfo {author} {\bibfnamefont
  {F.}~\bibnamefont {Caruso}}, \bibinfo {author} {\bibfnamefont {R.~E.}\
  \bibnamefont {Scholten}}, \ and\ \bibinfo {author} {\bibfnamefont {L.~C.~L.}\
  \bibnamefont {Hollenberg}},\ }\href {\doibase 10.1038/nnano.2011.64}
  {\bibfield  {journal} {\bibinfo  {journal} {Nat Nano}\ }\textbf {\bibinfo
  {volume} {6}},\ \bibinfo {pages} {358} (\bibinfo {year} {2011})}\BibitemShut
  {NoStop}%
\bibitem [{\citenamefont {Ramsey}(1990)}]{Ramsey90}%
  \BibitemOpen
  \bibfield  {author} {\bibinfo {author} {\bibfnamefont {N.~F.}\ \bibnamefont
  {Ramsey}},\ }\href@noop {} {\emph {\bibinfo {title} {Molecular Beams}}}\
  (\bibinfo  {publisher} {Oxford University Press},\ \bibinfo {year}
  {1990})\BibitemShut {NoStop}%
\bibitem [{\citenamefont {Hahn}(1950)}]{Hahn50}%
  \BibitemOpen
  \bibfield  {author} {\bibinfo {author} {\bibfnamefont {E.~L.}\ \bibnamefont
  {Hahn}},\ }\href {\doibase 10.1103/PhysRev.80.580} {\bibfield  {journal}
  {\bibinfo  {journal} {Phys. Rev.}\ }\textbf {\bibinfo {volume} {80}},\
  \bibinfo {pages} {580} (\bibinfo {year} {1950})}\BibitemShut {NoStop}%
\bibitem [{\citenamefont {Meiboom}\ and\ \citenamefont
  {Gill}(1958)}]{Meiboom58}%
  \BibitemOpen
  \bibfield  {author} {\bibinfo {author} {\bibfnamefont {S.}~\bibnamefont
  {Meiboom}}\ and\ \bibinfo {author} {\bibfnamefont {D.}~\bibnamefont {Gill}},\
  }\href@noop {} {\bibfield  {journal} {\bibinfo  {journal} {Rev. Sc. Instr.}\
  }\textbf {\bibinfo {volume} {29}},\ \bibinfo {pages} {688} (\bibinfo {year}
  {1958})}\BibitemShut {NoStop}%
\bibitem [{\citenamefont {Kosugi}\ \emph {et~al.}(2005)\citenamefont {Kosugi},
  \citenamefont {Matsuo}, \citenamefont {Konno},\ and\ \citenamefont
  {Hatakenaka}}]{Kosugi05}%
  \BibitemOpen
  \bibfield  {author} {\bibinfo {author} {\bibfnamefont {N.}~\bibnamefont
  {Kosugi}}, \bibinfo {author} {\bibfnamefont {S.}~\bibnamefont {Matsuo}},
  \bibinfo {author} {\bibfnamefont {K.}~\bibnamefont {Konno}}, \ and\ \bibinfo
  {author} {\bibfnamefont {N.}~\bibnamefont {Hatakenaka}},\ }\href {\doibase
  10.1103/PhysRevB.72.172509} {\bibfield  {journal} {\bibinfo  {journal} {Phys.
  Rev. B}\ }\textbf {\bibinfo {volume} {72}},\ \bibinfo {pages} {172509}
  (\bibinfo {year} {2005})}\BibitemShut {NoStop}%
\bibitem [{\citenamefont {Laraoui}\ and\ \citenamefont
  {Meriles}(2011)}]{Laraoui11}%
  \BibitemOpen
  \bibfield  {author} {\bibinfo {author} {\bibfnamefont {A.}~\bibnamefont
  {Laraoui}}\ and\ \bibinfo {author} {\bibfnamefont {C.~A.}\ \bibnamefont
  {Meriles}},\ }\href {\doibase 10.1103/PhysRevB.84.161403} {\bibfield
  {journal} {\bibinfo  {journal} {Phys. Rev. B}\ }\textbf {\bibinfo {volume}
  {84}},\ \bibinfo {pages} {161403} (\bibinfo {year} {2011})}; 
  \bibfield  {author} {\bibinfo {author} {\bibfnamefont {J.-M.}\ \bibnamefont
  {{Cai}}}, \bibinfo {author} {\bibfnamefont {B.}~\bibnamefont {{Naydenov}}},
  \bibinfo {author} {\bibfnamefont {R.}~\bibnamefont {{Pfeiffer}}}, \bibinfo
  {author} {\bibfnamefont {L.~P.}\ \bibnamefont {{McGuinness}}}, \bibinfo
  {author} {\bibfnamefont {K.~D.}\ \bibnamefont {{Jahnke}}}, \bibinfo {author}
  {\bibfnamefont {F.}~\bibnamefont {{Jelezko}}}, \bibinfo {author}
  {\bibfnamefont {M.~B.}\ \bibnamefont {{Plenio}}}, \ and\ \bibinfo {author}
  {\bibfnamefont {A.}~\bibnamefont {{Retzker}}},\ }\href
  {http://arxiv.org/abs/1111.0930} {\bibfield  {journal} {\bibinfo  {journal}
  {ArXiv}\ } (\bibinfo {year} {2011})},\ \Eprint
  {http://arxiv.org/abs/1111.0930} {arXiv:1111.0930};  
  \bibfield  {author} {\bibinfo {author} {\bibfnamefont {X.}~\bibnamefont
  {{Xu}}}, \bibinfo {author} {\bibfnamefont {Z.}~\bibnamefont {{Wang}}},
  \bibinfo {author} {\bibfnamefont {C.}~\bibnamefont {{Duan}}}, \bibinfo
  {author} {\bibfnamefont {P.}~\bibnamefont {{Huang}}}, \bibinfo {author}
  {\bibfnamefont {P.}~\bibnamefont {{Wang}}}, \bibinfo {author} {\bibfnamefont
  {Y.}~\bibnamefont {{Wang}}}, \bibinfo {author} {\bibfnamefont
  {N.}~\bibnamefont {{Xu}}}, \bibinfo {author} {\bibfnamefont {X.}~\bibnamefont
  {{Kong}}}, \bibinfo {author} {\bibfnamefont {F.}~\bibnamefont {{Shi}}},
  \bibinfo {author} {\bibfnamefont {X.}~\bibnamefont {{Rong}}}, \ and\ \bibinfo
  {author} {\bibfnamefont {J.}~\bibnamefont {{Du}}},\ }\href@noop {} {\bibfield
   {journal} {\bibinfo  {journal} {ArXiv e-prints}\ } (\bibinfo {year}
  {2012})},\ \Eprint {http://arxiv.org/abs/1205.1307} {arXiv:1205.1307}
  \BibitemShut {NoStop}%
\bibitem [{\citenamefont {Haeberlen}(1976)}]{Haeberlen76}%
  \BibitemOpen
  \bibfield  {author} {\bibinfo {author} {\bibfnamefont {U.}~\bibnamefont
  {Haeberlen}},\ }\href@noop {} {\emph {\bibinfo {title} {High Resolution NMR
  in Solids: Selective Averaging}}}\ (\bibinfo  {publisher} {Academic Press
  Inc., New York},\ \bibinfo {year} {1976})\BibitemShut {NoStop}%
\bibitem [{SOM()}]{SOM}%
  \BibitemOpen
  \href@noop {} {}\bibinfo {howpublished} {See supplementary online
  material.}\BibitemShut {Stop}%
\bibitem [{\citenamefont {Wineland}\ \emph {et~al.}(1992)\citenamefont
  {Wineland}, \citenamefont {Bollinger}, \citenamefont {Itano}, \citenamefont
  {Moore},\ and\ \citenamefont {Heinzen}}]{Wineland92}%
  \BibitemOpen
  \bibfield  {author} {\bibinfo {author} {\bibfnamefont {D.~J.}\ \bibnamefont
  {Wineland}}, \bibinfo {author} {\bibfnamefont {J.~J.}\ \bibnamefont
  {Bollinger}}, \bibinfo {author} {\bibfnamefont {W.~M.}\ \bibnamefont
  {Itano}}, \bibinfo {author} {\bibfnamefont {F.~L.}\ \bibnamefont {Moore}}, \
  and\ \bibinfo {author} {\bibfnamefont {D.~J.}\ \bibnamefont {Heinzen}},\
  }\href {\doibase 10.1103/PhysRevA.46.R6797} {\bibfield  {journal} {\bibinfo
  {journal} {Phys. Rev. A}\ }\textbf {\bibinfo {volume} {46}},\ \bibinfo
  {pages} {R6797} (\bibinfo {year} {1992})}\BibitemShut {NoStop}%
\bibitem [{\citenamefont {Balasubramanian}\ \emph {et~al.}(2009)\citenamefont
  {Balasubramanian}, \citenamefont {Neumann}, \citenamefont {Twitchen},
  \citenamefont {Markham}, \citenamefont {Kolesov}, \citenamefont {Mizuochi},
  \citenamefont {Isoya}, \citenamefont {Achard}, \citenamefont {Beck},
  \citenamefont {Tissler}, \citenamefont {Jacques}, \citenamefont {Hemmer},
  \citenamefont {Jelezko},\ and\ \citenamefont
  {Wrachtrup}}]{Balasubramanian09}%
  \BibitemOpen
  \bibfield  {author} {\bibinfo {author} {\bibfnamefont {G.}~\bibnamefont
  {Balasubramanian}}, \bibinfo {author} {\bibfnamefont {P.}~\bibnamefont
  {Neumann}}, \bibinfo {author} {\bibfnamefont {D.}~\bibnamefont {Twitchen}},
  \bibinfo {author} {\bibfnamefont {M.}~\bibnamefont {Markham}}, \bibinfo
  {author} {\bibfnamefont {R.}~\bibnamefont {Kolesov}}, \bibinfo {author}
  {\bibfnamefont {N.}~\bibnamefont {Mizuochi}}, \bibinfo {author}
  {\bibfnamefont {J.}~\bibnamefont {Isoya}}, \bibinfo {author} {\bibfnamefont
  {J.}~\bibnamefont {Achard}}, \bibinfo {author} {\bibfnamefont
  {J.}~\bibnamefont {Beck}}, \bibinfo {author} {\bibfnamefont {J.}~\bibnamefont
  {Tissler}}, \bibinfo {author} {\bibfnamefont {V.}~\bibnamefont {Jacques}},
  \bibinfo {author} {\bibfnamefont {P.~R.}\ \bibnamefont {Hemmer}}, \bibinfo
  {author} {\bibfnamefont {F.}~\bibnamefont {Jelezko}}, \ and\ \bibinfo
  {author} {\bibfnamefont {J.}~\bibnamefont {Wrachtrup}},\ }\href {\doibase
  10.1038/nmat2420} {\bibfield  {journal} {\bibinfo  {journal} {Nat Mater}\
  }\textbf {\bibinfo {volume} {8}},\ \bibinfo {pages} {383} (\bibinfo {year}
  {2009})}\BibitemShut {NoStop}%
\bibitem [{\citenamefont {Babinec}\ \emph {et~al.}(2010)\citenamefont
  {Babinec}, \citenamefont {M.}, \citenamefont {Khan}, \citenamefont {Zhang},
  \citenamefont {Maze}, \citenamefont {Hemmer},\ and\ \citenamefont
  {Loncar}}]{Babinec10}%
  \BibitemOpen
  \bibfield  {author} {\bibinfo {author} {\bibfnamefont {T.~M.}\ \bibnamefont
  {Babinec}}, \bibinfo {author} {\bibfnamefont {H.~J.}\ \bibnamefont {M.}},
  \bibinfo {author} {\bibfnamefont {M.}~\bibnamefont {Khan}}, \bibinfo {author}
  {\bibfnamefont {Y.}~\bibnamefont {Zhang}}, \bibinfo {author} {\bibfnamefont
  {J.~R.}\ \bibnamefont {Maze}}, \bibinfo {author} {\bibfnamefont {P.~R.}\
  \bibnamefont {Hemmer}}, \ and\ \bibinfo {author} {\bibfnamefont
  {M.}~\bibnamefont {Loncar}},\ }\href {\doibase 10.1038/nnano.2010.6}
  {\bibfield  {journal} {\bibinfo  {journal} {Nat Nano}\ }\textbf {\bibinfo
  {volume} {5}},\ \bibinfo {pages} {195} (\bibinfo {year} {2010})}\BibitemShut
  {NoStop}%
\bibitem [{\citenamefont {Neumann}\ \emph {et~al.}(2010)\citenamefont
  {Neumann}, \citenamefont {Beck}, \citenamefont {Steiner}, \citenamefont
  {Rempp}, \citenamefont {Fedder}, \citenamefont {Hemmer}, \citenamefont
  {Wrachtrup},\ and\ \citenamefont {Jelezko}}]{Neumann10b}%
  \BibitemOpen
  \bibfield  {author} {\bibinfo {author} {\bibfnamefont {P.}~\bibnamefont
  {Neumann}}, \bibinfo {author} {\bibfnamefont {J.}~\bibnamefont {Beck}},
  \bibinfo {author} {\bibfnamefont {M.}~\bibnamefont {Steiner}}, \bibinfo
  {author} {\bibfnamefont {F.}~\bibnamefont {Rempp}}, \bibinfo {author}
  {\bibfnamefont {H.}~\bibnamefont {Fedder}}, \bibinfo {author} {\bibfnamefont
  {P.~R.}\ \bibnamefont {Hemmer}}, \bibinfo {author} {\bibfnamefont
  {J.}~\bibnamefont {Wrachtrup}}, \ and\ \bibinfo {author} {\bibfnamefont
  {F.}~\bibnamefont {Jelezko}},\ }\href {\doibase 10.1126/science.1189075}
  {\bibfield  {journal} {\bibinfo  {journal} {Science}\ }\textbf {\bibinfo
  {volume} {5991}},\ \bibinfo {pages} {542} (\bibinfo {year}
  {2010})}\BibitemShut {NoStop}%
\bibitem [{\citenamefont {Klauder}\ and\ \citenamefont
  {Anderson}(1962)}]{Klauder62}%
  \BibitemOpen
  \bibfield  {author} {\bibinfo {author} {\bibfnamefont {J.~R.}\ \bibnamefont
  {Klauder}}\ and\ \bibinfo {author} {\bibfnamefont {P.~W.}\ \bibnamefont
  {Anderson}},\ }\href {\doibase 10.1103/PhysRev.125.912} {\bibfield  {journal}
  {\bibinfo  {journal} {Phys. Rev.}\ }\textbf {\bibinfo {volume} {125}},\
  \bibinfo {pages} {912} (\bibinfo {year} {1962})}\BibitemShut {NoStop}%
\bibitem [{\citenamefont {Hirose}\ \emph {et~al.}()\citenamefont {Hirose},
  \citenamefont {Aiello},\ and\ \citenamefont {Cappellaro}}]{Hirose12}%
  \BibitemOpen
  \bibfield  {author} {\bibinfo {author} {\bibfnamefont {M.}~\bibnamefont
  {Hirose}}, \bibinfo {author} {\bibfnamefont {C.~D.}\ \bibnamefont {Aiello}},
  \ and\ \bibinfo {author} {\bibfnamefont {P.}~\bibnamefont {Cappellaro}},\
  }\href@noop {} {}\bibinfo {note} {ArXiv:1207.5729v1}\BibitemShut {NoStop}%
\bibitem [{\citenamefont {Augustine}\ \emph {et~al.}(2003)\citenamefont
  {Augustine}, \citenamefont {Santamaria},\ and\ \citenamefont
  {Tanaka}}]{Augustine03}%
  \BibitemOpen
  \bibfield  {author} {\bibinfo {author} {\bibfnamefont {G.~J.}\ \bibnamefont
  {Augustine}}, \bibinfo {author} {\bibfnamefont {F.}~\bibnamefont
  {Santamaria}}, \ and\ \bibinfo {author} {\bibfnamefont {K.}~\bibnamefont
  {Tanaka}},\ }\href@noop {} {\bibfield  {journal} {\bibinfo  {journal}
  {Neuron}\ }\textbf {\bibinfo {volume} {40}},\ \bibinfo {pages} {331}
  (\bibinfo {year} {2003})}\BibitemShut {NoStop}%
\bibitem [{\citenamefont {Keller}\ \emph {et~al.}(2008)\citenamefont {Keller},
  \citenamefont {Franks}, \citenamefont {Bartol~Jr},\ and\ \citenamefont
  {Sejnowski}}]{Keller08}%
  \BibitemOpen
  \bibfield  {author} {\bibinfo {author} {\bibfnamefont {D.~X.}\ \bibnamefont
  {Keller}}, \bibinfo {author} {\bibfnamefont {K.~M.}\ \bibnamefont {Franks}},
  \bibinfo {author} {\bibfnamefont {T.~M.}\ \bibnamefont {Bartol~Jr}}, \ and\
  \bibinfo {author} {\bibfnamefont {T.~J.}\ \bibnamefont {Sejnowski}},\ }\href
  {\doibase 10.1371/journal.pone.0002045} {\bibfield  {journal} {\bibinfo
  {journal} {PLoS ONE}\ }\textbf {\bibinfo {volume} {3}},\ \bibinfo {pages}
  {e2045} (\bibinfo {year} {2008})}\BibitemShut {NoStop}%
\bibitem [{\citenamefont {Degen}(2008)}]{Degen08}%
  \BibitemOpen
  \bibfield  {author} {\bibinfo {author} {\bibfnamefont {C.~L.}\ \bibnamefont
  {Degen}},\ }\href {\doibase 10.1063/1.2943282} {\bibfield  {journal}
  {\bibinfo  {journal} {Applied Physics Letters}\ }\textbf {\bibinfo {volume}
  {92}},\ \bibinfo {eid} {243111} (\bibinfo {year} {2008})}\BibitemShut
  {NoStop}%
\bibitem [{\citenamefont {Grinolds}\ \emph {et~al.}(2011)\citenamefont
  {Grinolds}, \citenamefont {Maletinsky}, \citenamefont {Hong}, \citenamefont
  {Lukin}, \citenamefont {Walsworth},\ and\ \citenamefont
  {Yacoby}}]{Grinolds11}%
  \BibitemOpen
  \bibfield  {author} {\bibinfo {author} {\bibfnamefont {M.~S.}\ \bibnamefont
  {Grinolds}}, \bibinfo {author} {\bibfnamefont {P.}~\bibnamefont
  {Maletinsky}}, \bibinfo {author} {\bibfnamefont {S.}~\bibnamefont {Hong}},
  \bibinfo {author} {\bibfnamefont {M.~D.}\ \bibnamefont {Lukin}}, \bibinfo
  {author} {\bibfnamefont {R.~L.}\ \bibnamefont {Walsworth}}, \ and\ \bibinfo
  {author} {\bibfnamefont {A.}~\bibnamefont {Yacoby}},\ }\href {\doibase
  10.1038/nphys1999} {\bibfield  {journal} {\bibinfo  {journal} {Nat. Phys.}\
  }\textbf {\bibinfo {volume} {aop}},\  (\bibinfo {year} {2011})}\BibitemShut
  {NoStop}%
\bibitem [{\citenamefont {Wei}(2006)}]{Wei06}%
  \BibitemOpen
  \bibfield  {author} {\bibinfo {author} {\bibfnamefont {W.~W.~S.}\
  \bibnamefont {Wei}},\ }\href@noop {} {\emph {\bibinfo {title} {Time series
  analysis: Univariate and multivariate methods}}}\ (\bibinfo  {publisher}
  {Pearson Addison Wesley},\ \bibinfo {year} {2006})\BibitemShut {NoStop}%
\bibitem [{\citenamefont {Jacques}\ \emph {et~al.}(2009)\citenamefont
  {Jacques}, \citenamefont {Neumann}, \citenamefont {Beck}, \citenamefont
  {Markham}, \citenamefont {Twitchen}, \citenamefont {Meijer}, \citenamefont
  {Kaiser}, \citenamefont {Balasubramanian}, \citenamefont {Jelezko},\ and\
  \citenamefont {Wrachtrup}}]{Jacques09}%
  \BibitemOpen
  \bibfield  {author} {\bibinfo {author} {\bibfnamefont {V.}~\bibnamefont
  {Jacques}}, \bibinfo {author} {\bibfnamefont {P.}~\bibnamefont {Neumann}},
  \bibinfo {author} {\bibfnamefont {J.}~\bibnamefont {Beck}}, \bibinfo {author}
  {\bibfnamefont {M.}~\bibnamefont {Markham}}, \bibinfo {author} {\bibfnamefont
  {D.}~\bibnamefont {Twitchen}}, \bibinfo {author} {\bibfnamefont
  {J.}~\bibnamefont {Meijer}}, \bibinfo {author} {\bibfnamefont
  {F.}~\bibnamefont {Kaiser}}, \bibinfo {author} {\bibfnamefont
  {G.}~\bibnamefont {Balasubramanian}}, \bibinfo {author} {\bibfnamefont
  {F.}~\bibnamefont {Jelezko}}, \ and\ \bibinfo {author} {\bibfnamefont
  {J.}~\bibnamefont {Wrachtrup}},\ }\href {\doibase
  10.1103/PhysRevLett.102.057403} {\bibfield  {journal} {\bibinfo  {journal}
  {Phys. Rev. Lett.}\ }\textbf {\bibinfo {volume} {102}},\ \bibinfo {eid}
  {057403} (\bibinfo {year} {2009})}\BibitemShut {NoStop}%
\bibitem{Duer04}%
  \BibitemOpen
  \bibfield{author}{%
  \bibinfo {author} {\bibfnamefont{M.}~\bibnamefont{Duer}},\ }%
  \emph{\bibinfo {title} {Introduction to Solid-State NMR Spectroscopy}}\
  (\bibinfo {publisher} {John Wiley \& Sons},\ \bibinfo {year} {2004})%
  \BibitemShut {NoStop}
  \bibitem{Bretthorst88}%
  \BibitemOpen
  \bibfield{author}{%
  \bibinfo {author} {\bibfnamefont{G.~L.}\ \bibnamefont{Bretthorst}},\ }%
  \emph{\bibinfo {title} {Bayesian Spectrum Analysis and Parameter
  Estimation}}\ (\bibinfo {publisher} {Springer-Verlag},\ \bibinfo {year}
  {1988})%
  \BibitemShut {NoStop}
\bibitem{Rife74}%
  \BibitemOpen
  \bibfield{author}{%
  \bibinfo {author} {\bibfnamefont{D.~C.}\ \bibnamefont{Rife}}\ and\ \bibinfo
  {author} {\bibfnamefont{R.~R.}\ \bibnamefont{Boorstyn}},\ }%
  \bibfield{journal}{%
  \bibinfo {journal} {IEEE Transactions on Information Theory}\ }%
  \textbf{\bibinfo {volume} {20}},\ \bibinfo {pages} {591} (\bibinfo {year}
  {1974})%
  \BibitemShut {NoStop}
\bibitem{Wei06}%
  \BibitemOpen
  \bibfield{author}{%
  \bibinfo {author} {\bibfnamefont{W.~W.~S.}\ \bibnamefont{Wei}},\ }%
  \emph{\bibinfo {title} {Time series analysis: Univariate and multivariate
  methods}}\ (\bibinfo {publisher} {Pearson Addison Wesley},\ \bibinfo {year}
  {2006})%
  \BibitemShut {NoStop}
\bibitem{Meriles10}%
  \BibitemOpen
  \bibfield{author}{%
  \bibinfo {author} {\bibfnamefont{C.~A.}\ \bibnamefont{Meriles}}, \bibinfo
  {author} {\bibfnamefont{L.}~\bibnamefont{Jiang}}, \bibinfo {author}
  {\bibfnamefont{G.}~\bibnamefont{Goldstein}}, \bibinfo {author}
  {\bibfnamefont{J.~S.}\ \bibnamefont{Hodges}}, \bibinfo {author}
  {\bibfnamefont{J.}~\bibnamefont{Maze}}, \bibinfo {author}
  {\bibfnamefont{M.~D.}\ \bibnamefont{Lukin}},\ and\ \bibinfo {author}
  {\bibfnamefont{P.}~\bibnamefont{Cappellaro}},\ }%
  \bibfield{journal}{%
  {\bibinfo {journal} {J. Chem. Phys.}}\ }%
  \textbf{\bibinfo {volume} {133}},\ \bibinfo {pages} {124105} (\bibinfo {year}
  {2010})%
  \BibitemShut {NoStop}
\bibitem{Neumann10}%
  \BibitemOpen
  \bibfield{author}{%
  \bibinfo {author} {\bibfnamefont{P.}~\bibnamefont{Neumann}}, \bibinfo
  {author} {\bibfnamefont{R.}~\bibnamefont{Kolesov}}, \bibinfo {author}
  {\bibfnamefont{B.}~\bibnamefont{Naydenov}}, \bibinfo {author}
  {\bibfnamefont{J.}~\bibnamefont{Beck}}, \bibinfo {author}
  {\bibfnamefont{F.}~\bibnamefont{Rempp}}, \bibinfo {author}
  {\bibfnamefont{M.}~\bibnamefont{Steiner}}, \bibinfo {author}
  {\bibfnamefont{V.}~\bibnamefont{Jacques}}, \bibinfo {author}
  {\bibfnamefont{G.}~\bibnamefont{Balasubramanian}}, \bibinfo {author}
  {\bibfnamefont{M.~L.}\ \bibnamefont{Markham}}, \bibinfo {author}
  {\bibfnamefont{D.~J.}\ \bibnamefont{Twitchen}}, \bibinfo {author}
  {\bibfnamefont{S.}~\bibnamefont{Pezzagna}}, \bibinfo {author}
  {\bibfnamefont{J.}~\bibnamefont{Meijer}}, \bibinfo {author}
  {\bibfnamefont{J.}~\bibnamefont{Twamley}}, \bibinfo {author}
  {\bibfnamefont{F.}~\bibnamefont{Jelezko}},\ and\ \bibinfo {author}
  {\bibfnamefont{J.}~\bibnamefont{Wrachtrup}},\ }%
  \bibfield{journal}{%
  {\bibinfo {journal} {Nat Phys}}\ }%
  \textbf{\bibinfo {volume} {6}},\ \bibinfo {pages} {249} (\bibinfo {year}
  {2010})%
   \BibitemShut {NoStop}
\bibitem{Kubo62}%
  \BibitemOpen
  \bibfield{author}{%
  \bibinfo {author} {\bibfnamefont{R.}~\bibnamefont{Kubo}},\ }%
  \bibfield{journal}{%
  {\bibinfo {journal} {Journal of the Physical
  Society of Japan}}\ }%
  \textbf{\bibinfo {volume} {17}},\ \bibinfo {pages} {1100} (\bibinfo {year}
  {1962}),\ \url{http://jpsj.ipap.jp/link?JPSJ/17/1100/}%
   \BibitemShut {NoStop}
\bibitem{Dobrovitski09}%
  \BibitemOpen
  \bibfield{author}{%
  \bibinfo {author} {\bibfnamefont{V.~V.}\ \bibnamefont{Dobrovitski}}, \bibinfo
  {author} {\bibfnamefont{A.~E.}\ \bibnamefont{Feiguin}}, \bibinfo {author}
  {\bibfnamefont{R.}~\bibnamefont{Hanson}},\ and\ \bibinfo {author}
  {\bibfnamefont{D.~D.}\ \bibnamefont{Awschalom}},\ }%
  \bibfield{journal}{%
  \bibinfo {journal} {Phys. Rev. Lett.}\ }%
  \textbf{\bibinfo {volume} {102}},\ \bibinfo {pages} {237601} (\bibinfo {year}
  {2009})%
   \BibitemShut {NoStop}
\bibitem{Cappellaro06}%
  \BibitemOpen
  \bibfield{author}{%
  \bibinfo {author} {\bibfnamefont{P.}~\bibnamefont{Cappellaro}}, \bibinfo
  {author} {\bibfnamefont{J.~S.}\ \bibnamefont{Hodges}}, \bibinfo {author}
  {\bibfnamefont{T.~F.}\ \bibnamefont{Havel}},\ and\ \bibinfo {author}
  {\bibfnamefont{D.~G.}\ \bibnamefont{Cory}},\ }%
  \bibfield{journal}{%
  \bibinfo {journal} {J. Chem. Phys.}\ }%
  \textbf{\bibinfo {volume} {125}},\ \bibinfo {pages} {044514} (\bibinfo {year}
  {2006})%
   \BibitemShut {NoStop}
\bibitem{Khodjasteh05}%
  \BibitemOpen
  \bibfield{author}{%
  \bibinfo {author} {\bibfnamefont{K.}~\bibnamefont{Khodjasteh}}\ and\ \bibinfo
  {author} {\bibfnamefont{D.~A.}\ \bibnamefont{Lidar}},\ }%
  \bibfield{journal}{%
  {\bibinfo {journal} {Phys. Rev. Lett.}}\
  }%
  \textbf{\bibinfo {volume} {95}},\ \bibinfo {eid} {180501} (\bibinfo {year}
  {2005})%
   \BibitemShut {NoStop}
\bibitem{Jiang09}%
  \BibitemOpen
  \bibfield{author}{%
  \bibinfo {author} {\bibfnamefont{L.}~\bibnamefont{Jiang}}, \bibinfo {author}
  {\bibfnamefont{J.~S.}\ \bibnamefont{Hodges}}, \bibinfo {author}
  {\bibfnamefont{J.~R.}\ \bibnamefont{Maze}}, \bibinfo {author}
  {\bibfnamefont{P.}~\bibnamefont{Maurer}}, \bibinfo {author}
  {\bibfnamefont{J.~M.}\ \bibnamefont{Taylor}}, \bibinfo {author}
  {\bibfnamefont{D.~G.}\ \bibnamefont{Cory}}, \bibinfo {author}
  {\bibfnamefont{P.~R.}\ \bibnamefont{Hemmer}}, \bibinfo {author}
  {\bibfnamefont{R.~L.}\ \bibnamefont{Walsworth}}, \bibinfo {author}
  {\bibfnamefont{A.}~\bibnamefont{Yacoby}}, \bibinfo {author}
  {\bibfnamefont{A.~S.}\ \bibnamefont{Zibrov}},\ and\ \bibinfo {author}
  {\bibfnamefont{M.~D.}\ \bibnamefont{Lukin}},\ }%
  \bibfield{journal}{%
  {\bibinfo {journal} {Science}}\ }%
  \textbf{\bibinfo {volume} {326}},\ \bibinfo {pages} {267} (\bibinfo {year}
  {2009}),\ \url{http://www.sciencemag.org/cgi/content/abstract/326/5950/267}%
  \BibitemShut {NoStop}
\bibitem{Augustine03}%
  \BibitemOpen
  \bibfield{author}{%
  \bibinfo {author} {\bibfnamefont{G.~J.}\ \bibnamefont{Augustine}}, \bibinfo
  {author} {\bibfnamefont{F.}~\bibnamefont{Santamaria}},\ and\ \bibinfo
  {author} {\bibfnamefont{K.}~\bibnamefont{Tanaka}},\ }%
  \bibfield{journal}{%
  \bibinfo {journal} {Neuron}\ }%
  \textbf{\bibinfo {volume} {40}},\ \bibinfo {pages} {331} (\bibinfo {year}
  {2003})%
   \BibitemShut {NoStop}
\bibitem{Fakler08}%
  \BibitemOpen
  \bibfield{author}{%
  \bibinfo {author} {\bibfnamefont{B.}~\bibnamefont{Fakler}}\ and\ \bibinfo
  {author} {\bibfnamefont{J.~P.}\ \bibnamefont{Adelman}},\ }%
  \bibfield{journal}{%
  {\bibinfo {journal} {Neuron}}\ }%
  \textbf{\bibinfo {volume} {59}},\ \bibinfo {pages} {873} (\bibinfo {year}
  {2008}) \BibitemShut {NoStop}
\end{thebibliography}
\end{document}